\documentclass[a4paper,aps,prd,onecolumn,preprintnumbers,showpacs,nofootinbib]{revtex4}

\usepackage{amsmath,amssymb,graphics,epsfig,subfigure}
\usepackage{cases}
\usepackage{xcolor}
\usepackage[colorlinks,
            linkcolor=blue,
            anchorcolor=blue,
            citecolor=green
            ]{hyperref}
\usepackage[colorlinks,linkcolor=blue,anchorcolor=blue,citecolor=green]{hyperref}

\newcommand{\tcb}{\textcolor{blue}}

\begin{document}
\renewcommand{\baselinestretch}{1.3}

\title{Optical appearance of Einstein-\AE{}ther black hole surrounded by thin disk}

\author{Hui-Min Wang, Zi-Chao Lin,
        Shao-Wen Wei\footnote{weishw@lzu.edu.cn, corresponding author}}

\affiliation{Lanzhou Center for Theoretical Physics, Key Laboratory of Theoretical Physics of Gansu Province, School of Physical Science and Technology, Lanzhou University, Lanzhou 730000, People's Republic of China,\\
 Institute of Theoretical Physics $\&$ Research Center of Gravitation,
Lanzhou University, Lanzhou 730000, People's Republic of China,\\
 Academy of Plateau Science and Sustainability, Qinghai Normal University, Xining 810016, P. R. China}

\begin{abstract}
In Einstein-\AE ther theory, the Lorentz symmetry is locally broken in the high-energy regime due to the presence of the \AE ther field. This shall leave significant imprint on astronomical observation. In this paper, we investigate the optical appearance of two types of the static and spherically symmetric black holes in Einstein-\AE ther theory. Via Euler-Lagrange equation, we obtain the equations of motion of the photon and calculate the total deflection angle of the photon trajectory around the black hole. By classifying the light rays with the total number of orbits, we study the effects of coupling constants on the direct image, lensing ring, and photon ring. The features of the light trajectories are also investigated by comparing with the Einstein-\AE ther theory and general relativity. Moreover, we also show the explicit optical appearance of black holes surrounded by thin disk emissions with three characteristic emitted models. The results indicate that the direct image gives the main contribution to the total flux, and the lensing ring just gives a very small contribution, whereas the role of the photon ring is negligible. The optical appearances are also found to significantly rely on these coupling constants.
\end{abstract}


\maketitle

\section{Introduction}
\label{secIntroduction}

As a famous prediction in general relativity (GR), the black hole is a compact object that even the light could not escape from it. Schwarzschild quickly found the black hole solution, giving the first proof of the black hole theoretically. After that, the discovery of the neutron star observed by Bell gives a strong direct evidence for the existence of gravitationally collapsed compact objects. As a possible extremely compact object in the universe, black hole has then attracted special attentions in the fields of both astrophysics and theoretical physics. However, due to the limitation of observation techniques, people were unable to detect black holes directly, and can only predict or prove the existence of black holes through the indirect evidence. For example, Cygnus X-1, the scientific wager between Hawking and Thorne, was the first widely accepted source of black hole due to its intense X-ray emission~\cite{BLWebster,CTBolton}. Another interesting example is that from observations of several stars orbiting the black hole, particularly star S2, astronomers have determined that Sagittarius A$^*$, located at the heart of the Milky Way galaxy, contains a supermassive black hole of about 4.3 million solar masses~\cite{RSchodel}. It was not until 11 February 2016, that LIGO and Virgo collaboration announced the first detection of gravitational waves generated by a binary black hole merger~\cite{BPAbbott}. It breaks the situation that black holes could only be explored indirectly. Excitingly, on 10 April 2019, the Event Horizon Telescope (EHT) collaboration released the first image of the supermassive black hole in the center of the giant elliptical galaxy M87 and its vicinity~\cite{KAkiyama1,KAkiyama2,KAkiyama3,KAkiyama4,
KAkiyama5,KAkiyama6}. This is the first time that humans directly observe the black hole. The image shows a dark region surrounded by a bright ring, called the critical curve, which is made up of photons trapped because of the gravitational bending. The dark region in the middle is the black hole shadow, which successfully confirms the prediction of GR.

The optical appearance of a black hole depends on the trajectory of surrounding photons, which in turn depends on the geometry of the spacetime. As early as 1966, the escape of photons from the surface of gravitationally intense stars was studied by Synge~\cite{JLSynge}. After that Luminet considered the image of a spherical black hole with a thin accretion disk~\cite{JPLuminet}. For a Schwarzschild black hole with mass $M$, the bounded photon orbit has the radius of $r_p=3M$, and the value of the critical impact parameter is $b_c=3\sqrt{3}M$. Besides, the shadow of such a non-rotating black hole is a standard circle. However, as was first studied by Bardeen in Refs.~\cite{JMBardeen1,JMBardeen2,JMBardeen3}, the black hole shadow of a rotating black hole may have more fine structures. It was later found that although the shadow of a Kerr black hole deviates from the circular shape, its D-shaped shadow, which evolves with the spin parameter, still has approximately the same typical radius of the critical curve~\cite{HFalcke,TJohannsen}. In order to fit the theoretical model by the astronomical observation, various observables of the shadow are proposed. For example, the radius $R_{\rm s}$ of the reference circle showing the shadow size, as well as the distortion parameter $\delta_{\rm s}$, which describes the degree of deformation between the shadow and the reference circle, were introduced by Hioki and Maeda~\cite{KHioki}. To eliminate the degeneracy, more distortion parameters are necessary~\cite{NTsukamoto,RKumar}. So far, there have been a lot of works focusing on the shadow cast by different black holes. In Ref.~\cite{CBambi}, Bambi and Freese considered super-spinning black holes ($J>M$), which violate the Kerr bound. They found that observers on the equatorial plane would see fan-shaped shadows, while those on the non-equatorial plane would see elliptical shadows. Besides, it is very difficult for astronomical observations to pinpoint the location of the center of shadow. To solve this problem, a general, coordinate-independent formalism was proposed~\cite{AAbdujabbarov}. In this formalism, shadow is described as an arbitrary polar curve represented by a Legendre expansion. More interestingly, in Ref. \cite{SWWei1}, we proposed the concept of the curvature radius for the black hole shadow, which is also independent of the location of the center of shadow, while is dependent of the intrinsic properties of the critical curve. On the other hand, the Schwarzschild black hole shadows in an external gravitational field~\cite{SAbdolrahimi}, with a cosmological constant~\cite{GSBisnovatyi1} and with halos~\cite{MWang} were well studied. Kerr black hole shadows in different backgrounds have also been extensively investigated~\cite{PVPCunha1,
HMWang1,SWWei1,SWWei2,PVPCunha3,HMWang2,SWWei3}. In addition, works on other types of black hole shadow are also very interesting, such as a near-extremal Kerr-Sen black hole~\cite{MGuo}, Kerr-Taub-NUT black hole~\cite{MZhang}, and so on~\cite{PVPCunha2,LMa,OYTsupko,PVPCunha4,SWWei4,JSolanki,
RCPantig,QMFu}.

In 2019, a very interesting viewpoint was put forward by Gralla, Holz, and Wald in Ref.~\cite{SEGralla}, which classifies the photons into three categories according to the total number of orbits rotating around the black hole. The image formed by the photons with the total number of orbits less than 3/4 is called the direct image, the image with the number of orbits between 3/4 and 5/4 is the lensing ring, and the image with the number larger than 5/4 is the photon ring. This novel perspective provides a new way to study the black hole image. Taking advantage of this method, a black hole in the quintessence dark energy model was investigated~\cite{XXZeng1}. It found that the quintessence state parameter has an influence on the distance between the observer and the event horizon, and thus affects the shadow image. When quantum corrections are taken into account, it turns out that the quantum corrected Schwarzschild black hole has a larger bright lensing ring than that of the classical Schwarzschild one~\cite{JPeng}. In addition, $f(R)$ global monopole black hole~\cite{GPLi1}, Kehagias-Sfetsos black hole~\cite{GPLi2}, black bounces~\cite{MGuerrero}, noncommutative black hole~\cite{XXZeng2}, Brane-World black hole~\cite{XXZeng3}, and black hole with torsion charge~\cite{KJHe}, etc. have been studied by this method. And their results all show that, the direct image always dominates the total flux, and the lensing ring just gives a very small contribution, whereas the photon ring makes a negligible contribution.

On the other hand, considering an ideal scenario where the photons are all originated from a source at infinity, the black hole shadow seen by the observer at infinity is the dark region inside the critical curve. And in this dark region, there is no other image with brightness. However, astrophysical black holes are always surrounded by a massive amount of accretion material. And these accretion materials may interact with photons from infinity or emit photons themselves. Thus they will affect the optical appearance of the black hole. According to it, many accretion models have been proposed, such as the thin disk accretion~\cite{JPLuminet,SEGralla,CLiu1,GSBisnovatyi2}, geometrically thick accretion~\cite{SEGralla}, spherically symmetric accretion~\cite{RNarayan,KSaurabh,XQin,QGan}, heavy accretion~\cite{PVPCunha5}, and so on. We should note that, with the accretion material extending to the event horizon, the dark region of the black hole image will be confined into a region, called the inner shadow, smaller than the black hole shadow. It means that the image of the black hole will still be visible inside the shadow~\cite{AChael}. And it seems useful to reveal the new physics in the high-energy regime of a black hole.

As is well-known that GR does not take into account the quantum effects, it then preserves the Lorentz invariance in the theory. However, when the quantization of the spacetime is considered in high-energy regime, Lorentz invariance could break down~\cite{DMattingly}. Therefore, the test of the Lorentz violation is a challenge to our fundamental physics. This spirit has been applied to many modified gravitational theories, such as the loop quantum gravity~\cite{AAshtekar}, Einstein-bumblebee gravity~\cite{VAKostelecky}, and Einstein-\AE ther gravity~\cite{TJacobson1,SMCarroll,CEling1,TJacobson2,TJacobson3}. Among them, Einstein-\AE ther gravity attracts much more attention, recently. In the framework of Einstein-\AE ther theory, there are three gravitational modes, the scalar (spin-0 graviton), vector (spin-1 graviton), and tensor (spin-2 graviton), respectively. Their speeds depend on four coupling constants. It is worth noting that when the local spherically symmetric spacetime is considered, the spin-1 and spin-2 gravitons are not excited, while only the spin-0 graviton works. Moreover, in order to guarantee the Lorentz violation locally, a unit time-like vector field, the \AE ther field, is introduced in the theory. As early as 2013, the constraints on Einstein-\AE ther theory from binary pulsar observations were studied~\cite{KYagi}. Later, according to the observations on GW 170817 and GRB 170817A, the experimental constraints on the coupling constants were given in Refs.~\cite{YGong,JOost} respectively. Besides, some new constraints were discussed based on the data of M87$^*$ from the EHT observations~\cite{MKhodadi}.

Afterwards, black holes with the presence of the \AE ther field have attracted more and more attention. After studying Einstein-\AE ther theory, Jacobson \textit{et al.} turned their attentions to black hole solutions in GR coupled to the \AE ther field~\cite{CEling2}. It is found that within a wide range of couplings external to the event horizon, these solutions are very close to their Schwarzschild counterparts. And inside the event horizon, there is a space-like singularity, but some differences in physical quantities exist. After a short while, in the non-reduced Einstein-\AE ther theory, the evolution of the gravitational perturbations of a spherically symmetric black hole, both in frequency and time domains was considered by Konoplya and Zhidenko~\cite{RAKonoplya}. The results show that compared with Schwarzschild black hole, the actual oscillation frequency and damping rate of Einstein-\AE ther black holes are larger. Besides, not far inside the metric horizon, these black hole solutions have a universal horizon that captures waves of any speed~\cite{EBarausse1}. Berglund \textit{et al.} pointed that the universal horizon still obeys the first law of black hole thermodynamics, and if it also obeys the second law, it will have entropy~\cite{PBerglund1,PBerglund2}. By coupling a scalar field with the time-like vector, they concluded that even though the scalar field equations violate the local Lorentz invariance, the universal horizon still radiates acting as a black body. In addition, two new types of exact charged black hole solutions in Einstein-\AE ther theory were found in Ref.~\cite{CDing}. Slowly rotating black holes were also studied~\cite{EBarausse2}, and unconstrained by the naked finite area singularities, the solutions form a two-parameter family, which can be regarded as the mass and angular momentum of the black hole. There is also a lot of research on black holes in the framework of this theory~\cite{TZhu,CZhang,JRayimbaev,CLiu,AAdam}. In Ref.~\cite{TZhu}, the black hole shadow in the traditional manner was studied. It was found that different coupling constants describing the \AE ther field have different effects on the size of the black hole shadow. In this paper, we will show how the \AE ther field contributes to the inner shadow of a static and spherically symmetric black hole when the thin disk is present.

This paper is organized as follows. In Sec.~\ref{theory}, we briefly review Einstein-\AE ther gravity and two types of exact black hole solutions. Employing the Euler-Lagrange equation, we study the motion of photons in Sec.~\ref{photonmotion}. According to the total number of orbits, we examine the direct image, lensing ring, and photon ring. In Sec.~\ref{image}, three toy-model functions are used to study the emitted specific intensity. The observed intensity is also investigated. Then we delineate the images of black hole surrounded by thin disk emissions. Finally, we summarize our results in Sec.~\ref{conclusions}.

\section{Einstein-\AE ther theory and the static spherically symmetric black hole solutions}
\label{theory}
In Einstein-\AE ther theory, the Lorentz symmetry is locally broken due to the presence of the \AE ther field. The action of the theory reads~\cite{TJacobson2}
\begin{eqnarray}
  S_{\ae}=\frac{1}{16\pi G_{\ae}}\int d^{4}x \sqrt{-g}(R+\mathcal{L}_{\ae}).\label{eq:action}
\end{eqnarray}
In this action, $G_{\ae}$ is the \AE ther gravitational constant, $g$ is the determinant of the spacetime metric $g_{\mu\nu}$ with the signature ($-$,+,+,+), and $R$ is the Ricci scalar. Besides, the Lagrangian of the \AE ther field is given by
\begin{eqnarray}\label{actionae}
  \mathcal{L}_{\ae}\equiv -(c_1g^{\alpha\beta}g_{\mu\nu}+c_2\delta^\alpha _\mu \delta^\beta_\nu+c_3\delta^\alpha_\nu \delta^\beta_\mu-c_4u^\alpha u^\beta g_{\mu\nu})(\nabla_\alpha u^\mu)(\nabla_\beta u^\nu)+\lambda(g_{\mu\nu}u^\mu u^\nu+1),
\end{eqnarray}
where $\lambda$ acts as a Lagrangian multiplier, and $u^\mu$ denotes the four-velocity of the \AE ther field, which is always time-like and unity guaranteed by $\lambda$. Moreover, we notice that there are four dimensionless coupling constants ($c_1,c_2,c_3,c_4$). Terms involving these coupling constants give the dynamics of the \AE ther vector field, and will locally break the Lorentz symmetric at high energy. So they are related to the new physics beyond the standard model in the high-energy regime. The Newtonian constant $G_N$ is related with the \AE ther gravitational constant $G_{\ae}$ via the coupling constants $c_1$ and $c_4$~\cite{SMCarroll}:
\begin{eqnarray}
  G_{\ae}=\frac{2G_N}{2-c_1-c_4}.
\end{eqnarray}
As a result, we can see that the coupling constants $c_1$ and $c_4$ determine the nature of gravity. By adjusting them, an anti-gravity with negative $G_{\ae}$ can be realized.

The variations of the action with respect to $g_{\mu\nu}$, $u^\mu$, and $\lambda$ yield, respectively, the equations of field
\begin{eqnarray}
R^{\mu\nu}-\frac{1}{2}g^{\mu\nu}R-8\pi G_{\ae}T^{\mu\nu}_{\ae}&=&0, \label{eq:fieldeq1} \\
\nabla_\mu {J^\mu}_\alpha +c_4a_\mu \nabla_\alpha u^\mu+\lambda u_\alpha&=&0,\label{eq:fieldeq2} \\
g_{\mu\nu}u^\mu u^\nu&=&-1 \label{eq:fieldeq3},
\end{eqnarray}
where
\begin{eqnarray}
T_{\alpha\beta}^{\ae} &\equiv& \nabla_\mu \left({J^\mu}_{(\alpha} u_{\beta)}+J_{(\alpha \beta)}u^\mu-u_{(\beta}{J_{\alpha)}}^\mu\right)+ c_1\Big(\big(\nabla_{\alpha}u_{\mu}\big)\big(\nabla_{\beta}
u^{\mu}\big) - \big(\nabla_{\mu}u_{\alpha}\big)\big(\nabla^{\mu}u_{\beta}
\big)\Big)\nonumber \\
&& + c_4 a_{\alpha}a_{\beta} + \lambda  u_{\alpha}u_{\beta} - \frac{1}{2} g_{\alpha\beta} {J^{\delta}}_{\sigma} \nabla_{\delta}u^{\sigma},\nonumber \\
{J^\alpha}_\mu &\equiv& (c_1g^{\alpha\beta}g_{\mu\nu}+c_2\delta^\alpha _\mu \delta^\beta_\nu+c_3\delta^\alpha_\nu \delta^\beta_\mu-c_4u^\alpha u^\beta g_{\mu\nu})\nabla_\beta u^\nu,\nonumber \\
a^\mu &\equiv& u^\alpha \nabla_\alpha u^\mu.
\end{eqnarray}
We should note that the field equation~\eqref{eq:fieldeq3} just imposes a time-like condition to the \AE ther vector field $u^{\mu}$ through the last term in Eq.~\eqref{actionae}.
Moreover, Eqs.~\eqref{eq:fieldeq2} and \eqref{eq:fieldeq3} show us that
\begin{eqnarray}
\lambda=u_\beta \nabla_\alpha J^{\alpha \beta}+c_4a_\lambda a^\lambda.
\end{eqnarray}
In this paper, we will focus on the static and spherically symmetric black hole solutions with the following metric ansatz
\begin{eqnarray}
  \text{d}s^2 = -f(r) \text{d}t^2 + f^{-1}(r)\text{d}r^2 + r^2 (\text{d}\theta^2 + \sin^2 \theta \text{d}\phi^2).\label{eq:ds}
\end{eqnarray}
On the boundary, we require that it reduces to the Minkowski metric. The corresponding Killing vector $\chi^\mu$ is given by
\begin{eqnarray}
\chi^\mu=(1,0,0,0),
\end{eqnarray}
and the \AE ther vector field $u^\mu$ is assumed as
\begin{eqnarray}
u^\mu(r)=\Big(\alpha(r)-\frac{\beta(r)}{f(r)},\beta(r),0,0\Big).
\end{eqnarray}
In order to coincide with the boundary condition, one requires $u^\mu(r\rightarrow +\infty)=(1,0,0,0)$.

As shown in Refs.~\cite{CEling2,EBarausse1,PBerglund2}, two kinds exact black hole solutions were obtained by solving these equations of field. Both them highly depend on these coupling constants. In the first Einstein-\AE{}ther black hole solution, the coupling constants satisfy $c_{14}\equiv c_{1}+c_{4}=0$ and $c_{123}\equiv c_{1}+c_{2}+c_{3}\neq0$. The metric function $f(r)$ can be expressed as
\begin{eqnarray}
  f_1(r) = 1 -\frac{2M}{r}-\frac{27c_{13}}{256(1-c_{13})} \left( \frac{2M}{r}\right)^4,\label{eq:f1}
\end{eqnarray}
where $c_{13}\equiv c_{1}+c_{3}$. The functions $\alpha(r)$ and $\beta(r)$ of the \AE ther vector field are given by~\cite{PBerglund2}
\begin{eqnarray}
\alpha_1(r)&=&\left(\frac{3\sqrt{3}}{16\sqrt{1-c_{13}}}
\Big(\frac{2M}{r}\Big)^2+\sqrt{1 -\frac{2M}{r}+\frac{27}{256} \Big( \frac{2M}{r}\Big)^4}\right)^{-1},\\
\beta_1(r)&=&-\frac{3\sqrt{3}}{16\sqrt{1-c_{13}}}
\left(\frac{2M}{r}\right)^2.
\end{eqnarray}
Taking $c_{13}\rightarrow0$, Schwarzschild black hole solution will be recovered. The radius of the event horizon of the first Einstein-\AE ther black hole solution~\eqref{eq:f1} can be obtained by solving $f_1(r)=0$, which gives
\begin{eqnarray}
  r_{01}=\frac{1}{4}\left( 2M+a_4+2\sqrt{2M^2-\frac{a_2}{4}-\frac{3a_3}{2}+\frac{4M^3}{a_4}}   \right),\label{eq:r01}
\end{eqnarray}
with
\begin{subequations}
  \begin{eqnarray}
	 a_1\!&\!=\!&\!\left( M^6c_{13}-2M^6c_{13}^2+M^6c_{13}^3+\sqrt{-M^{12}(-1+c_{13})^3c_{13}^2}\right)^{1/3}, \\
	  a_2\!&\!=\!&\!\frac{6M^4c_{13}}{a_1},\quad a_3=\frac{a_1}{-1+c_{13}},
	  \quad
	  a_4=\sqrt{4M^2+a_2+6a_3}\,.
  \end{eqnarray}
\end{subequations}
Taking $c_{123}=0$, the second Einstein-\AE ther solution is given by
\begin{eqnarray}
  f_2(r) = 1 -\frac{2M}{r}+\frac{M^2(c_{14}-2c_{13})}{2r^2(1-c_{13})},\label{eq:f2}
\end{eqnarray}
together with the solutions of the \AE ther vector field
\begin{eqnarray}
\alpha_2(r)&=&\left( 1+\frac{M}{r}\Big(\sqrt{\frac{2-c_{14}}{2(1-c_{13})}}-1 \Big) \right)^{-1},\\
\beta_2(r)&=&-\frac{M}{r}\sqrt{\frac{2-c_{14}}{2(1-c_{13})}}\,.
\end{eqnarray}
Obviously, one gets Schwarzschild solution when $c_{13}=c_{14}=0$. For this black hole solution, the radius of the event horizon is
\begin{eqnarray}
  r_{02}=M\Bigg(1-\sqrt{\frac{c_{14}-2}{2(c_{13}-1)}}\,\Bigg).\label{eq:r02}
\end{eqnarray}
So far, a number of constraints on the coupling constants are given by combining the theoretical and observational results~\cite{TJacobson2,TJacobson3,MKhodadi,KYagi,YGong,JOost}. In this paper, in order to see the effect of these coupling constants on the black hole shadow more clearly, we will adopt the following constraints \cite{TJacobson2,TJacobson3}:
\begin{eqnarray}
  c_{13}<1,\quad 0\leq c_{14}<2,\quad  2+c_{13}+3c_2>0.
\end{eqnarray}

\section{Null geodesic in Einstein-\AE ther spacetime}
\label{photonmotion}

It is worth noting that, black holes cannot be observed directly due to their strong gravitational pull. However, we can explore the black holes via their astronomical observable phenomena around them. It is well known that the geometry of the black hole spacetime determines the trajectory of surrounding photons, and many phenomena are formed by these photons, so one of the best and most mature ways to understand the black holes is investigating the motion of the photon. For this purpose, in this section, we will study the motion of the photon orbiting the black holes by the Euler-Lagrange approach.

Reminding the spherical symmetry of the black hole solution~\eqref{eq:ds}, we will focus on the motion of the photon confined on the equatorial plane with $\theta=\pi/2$ for convenience. Then the Lagrangian $\mathcal{L}$ for the photon is given by
\begin{eqnarray}
  \mathcal{L}=\frac{1}{2}g_{\mu\nu}\dot{x}^\mu\dot{x}^\nu
  =\frac{1}{2}\left( -f(r)\dot{t}^2+\frac{\dot{r}^2}{f(r)}+r^2{\dot{\phi}^2} \right),\label{eq:lag}
\end{eqnarray}
where $\dot{x}^\mu$ is the four-velocity, and thereafter, we will use dots to denote the derivative with respect to the affine parameter \tcb{$\tau$}. The explicit forms of the motion of the photon can be obtained by solving the Euler-Lagrange equation
\begin{eqnarray}
  \frac{d}{d\tau}\left( \frac{\partial \mathcal{L} }{\partial \dot{x}^\mu}\right)
  =\frac{\partial \mathcal{L} }{\partial x^\mu}.\label{eq:elag}
\end{eqnarray}
Since the coordinates $t$ and $\phi$ are absent in the Lagrangian $\mathcal{L}$ explicitly, we can get two conserved quantities, the energy $E$ and angular momentum $L$:
\begin{eqnarray}
  E=-\frac{\partial \mathcal{L}}{\partial \dot{t}}=f(r)\dot{t}, \quad
  L=\frac{\partial \mathcal{L}}{\partial \dot{\phi}}=r^2\dot{\phi}.\label{eq:energy}
\end{eqnarray}
Now, we turn to the radial motion of the photon. Taking advantage of $\mathcal{L}=0$ for a photon, we obtain
\begin{eqnarray}
  \dot{r}=\sqrt{E^2-\frac{L^2}{r^2}f(r)}\,,\label{eq:dotr}
\end{eqnarray}
where we have used Eqs.~\eqref{eq:lag} and \eqref{eq:energy}. By introducing the impact parameter $b\equiv L/E$, one can get the motion equation of the photon in the background of Einstein-\AE ther black hole
\begin{eqnarray}
  R(r)=\frac{\dot{r}^{2}}{L^{2}}=\frac{1}{b^2}-\frac{f(r)}{r^2}.
\end{eqnarray}
Obviously, $R(r)$ should be non-negative for a photon with certain values of $b$. The conditions determining the circular null orbits are given by
\begin{equation}
  R(r)=0,\quad R'(r)=0,
\end{equation}
where the prime denotes the derivative with respect to $r$. Solving them, we can acquire the radius of the photon sphere $r_p$ and the corresponding critical impact parameter $b_c$. Through the equations of motion, when the impact parameter of a photon is less than $b_c$, it will gradually approach and fall into the black hole. This kind of photon cannot escape from the black hole and will never reach the observer. As a result, any photon within the critical impact parameter $b_c$ will fall into the black hole, and form an unobtainable dark region known as the black hole shadow. And the actual radius of this region is related by $r_p$. Conversely, when the impact parameter of a photon is greater than $b_c$, it will escape from the black hole and eventually reach the observer at infinity. It gives the radius of the black hole shadow, which is always greater than $r_p$. For the first Einstein-\AE{}ther black hole solution, the two quantities, $r_p$ and $b_c$, have complicated forms, and we will not show them here. While for the second Einstein-\AE{}ther black hole solution, we have
\begin{subequations}
  \begin{eqnarray}
		r_{p2}&=&\frac{1}{2}M\left( 3+\sqrt{\frac{-9+c_{13}+4c_{14}}{-1+c_{13}}}\right),\label{eq:rp2}
		\\
		b_{c2}&=&M\sqrt{\frac{27-c_{13}^2-9a_5+c_{13}
		(-18+10c_{14}+a_5)+2c_{14}(-9+c_{14}+2a_5)}
		{(-1+c_{13})(-2+c_{14})}}\,,\label{eq:bc2}
  \end{eqnarray}
\end{subequations}
where $a_5=\sqrt{-1+c_{13}}\sqrt{-9+c_{13}+4c_{14}}$.

On the other hand, by making use of Eqs.~\eqref{eq:energy} and \eqref{eq:dotr}, we can rewrite the orbit equation of the photon as follows:
\begin{eqnarray}
  \frac{dr}{d\phi}=\sqrt{\frac{r^4}{b^2}-f(r)r^2}.
\end{eqnarray}
For the convenience of the following calculation, we make a coordinate transformation $u=1/r$. For the first kind of the black hole solution with $c_{14}=0$ and $c_{123}\neq0$, we have
\begin{equation}
  \Phi_1(u)\equiv\frac{\text{d}u}{\text{d}\phi} =\sqrt{\frac{1}{b^2}+u^2(-1+2Mu)+\frac{27M^4u^6c_{13}}{16-16c_{13}}}\,,\label{eq:phiu1}
\end{equation}
through which the trajectory of a photon in the equatorial plane can be obtained. Supposing that a photon with the impact parameter $b$ approaches the black hole from infinity, we have the following three situations:

(1) $b>b_c$. The photon gradually approaches the black hole along a curved trajectory, when the closest point $u_{\text{min}}$ is reached, it will escape from the black hole, and eventually return to the infinity. The turning point exactly corresponds to the minimum positive real root of $\Phi_1(u)=0$. And due to the symmetry of the photon trajectory, the total deflection angle of the trajectory can be obtained from the following equation
\begin{equation}
  \phi=2\int^{u_{\text{min}}}_0\!\!\frac{\text{d}u}
  {\sqrt{\Phi_1(u)}}.\label{eq:phi1}
\end{equation}

(2) $b=b_c$. It gives a class of photons in the critical case. Similar to the first case, the trajectory will be bent towards the black hole until the photon reaches the critical point $u_c$. Then the photon will not escape to the infinity, while revolve around the black hole along a circular orbit with radius $r_{p}$ instead. Therefore, such photons form the boundary of the black hole shadow.

(3) $b<b_c$. Before falling into the black hole, most of this kind of photons will also approach the black hole along a curved trajectory, just like these photons in the two previous cases. The difference is that they can neither escape from the black hole, nor circle around the black hole with enough angular momentum. So they will unavoidably get closer and closer to the event horizon (thereafter we set the radius of the event horizon as $r_0=1/u_0$). Here we should mention that, for most of these photons, their trajectories are curved, while for the rest photons that travel directly toward the black hole, the trajectories are straight lines. As a result, photons with $b<b_c$ will eventually fall into the black hole. Hence the total deflection angle is
\begin{equation}
  \phi=\int^{u_0}_0 \frac{du}{\sqrt{\Phi_1(u)}}.\label{eq:phi2}
\end{equation}
From the above analyses, we can obtain the deflection angle for these photons with different values of the impact parameter $b$.

Following the novel viewpoints of Ref.~\cite{SEGralla}, we will study the black hole image cast by the Einstein-\AE ther black holes. It is convenient to introduce the quantity  $n\equiv \phi/(2\pi)$, which measures the total number of the loops made by the photon around the black hole. Obviously, since the deflection angle depends on the impact parameter, $n$ is a function of $b$ as expected. Before carrying out the specific calculation, we would like to describe the configuration. For a fixed equatorial plane of the black hole, we place the observer at the north pole. The light source is an infinite plane located in the opposite side at infinity. According to the value of $n$, one could classify these light rays into three categories: (1) Direct image: $n<3/4$, (2) Lensing ring: $3/4<n<5/4$, and (3) Photon ring: $n>5/4$. These light rays intersect with the equatorial plane at most one time, twice times, and at least three times, respectively.

\begin{figure*}
\begin{center}
\subfigure[]{\label{bnf1all}
\includegraphics[width=5.5cm]{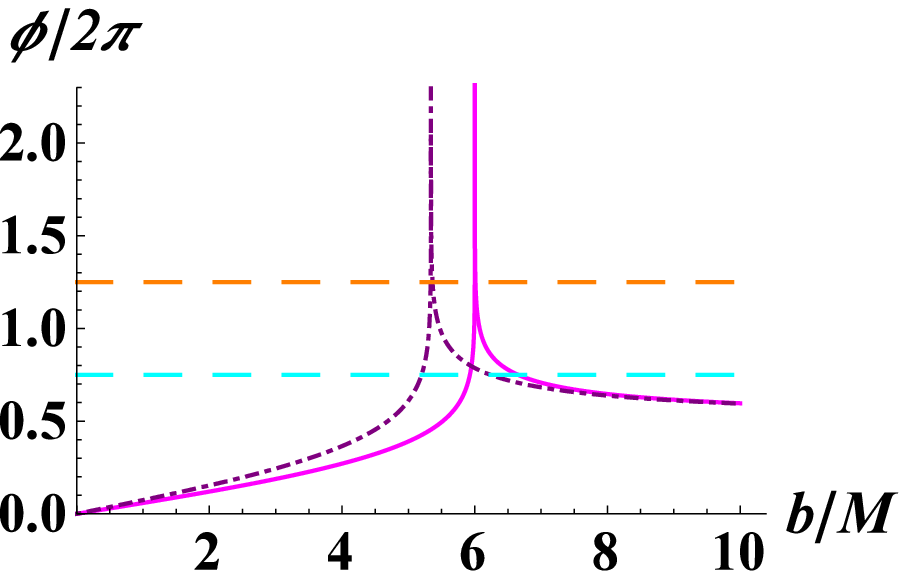}}
\quad
\subfigure[]{\label{bnf21all}
\includegraphics[width=5.5cm]{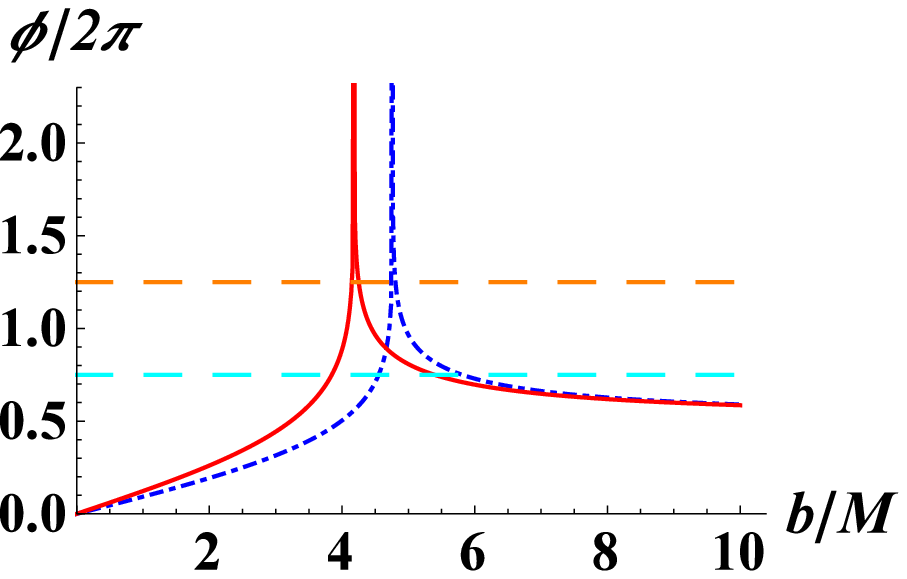}}
\quad
\subfigure[]{\label{bnf22all}
\includegraphics[width=5.5cm]{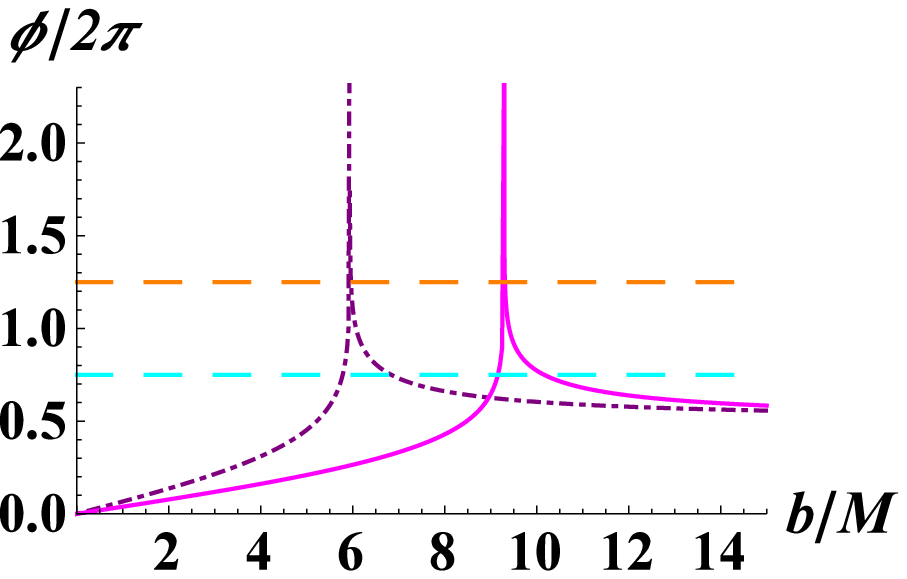}}
\caption{The deflection angle of photons in Einstein-\AE ther black hole as a function of the impact parameter $b/M$. The cyan and orange dashed lines represent $n=3/4$ and $n=5/4$, respectively. (a) For the first Einstein-\AE{}ther black hole solution with $c_{14}=0$ and $c_{123}\neq0$. $c_{13}=0.5$ (purple dot dashed curve) and $c_{13}=0.9$ (magenta curve). (b) For the second Einstein-\AE{}ther black hole solution with $c_{123}=0$. $c_{13}=0$ and $c_{14}=0.9$ (blue dot dashed curve), $c_{13}=0$ and $c_{14}=1.8$ (red curve). (c) For the second Einstein-\AE{}ther black hole solution with $c_{123}=0$. $c_{14}=0$ and $c_{13}=0.5$ (purple dot dashed curve), $c_{14}=0$ and $c_{13}=0.9$ (magenta curve).}\label{fig1}
\end{center}
\end{figure*}

For the first Einstein-\AE ther black hole solution with $c_{14}=0$ and $c_{123}\neq0$, we plot the total number of orbits $n$ as a function of $b/M$ in Fig.~\ref{bnf1all}. The cyan and orange dashed horizontal lines correspond to $n=3/4$ and $n=5/4$, respectively. For the purple dot dashed curve, we set $c_{13}=0.5$. With the increase of the impact parameter $b/M$, $n$ increases accompanied by an increasing slope, and reaches a peak at $b/M\approx 5.3$. Then it rapidly decreases to $n=5/4$, and finally decreases slowly and gradually approaches $n=1/2$. This pattern indicates that as $b/M$ increases, the deflection of light rays becomes less apparent, and the photon trajectory gradually tends to a straight line. Besides, our results show that the direct image with $n<3/4$ corresponds to a large range of $b/M$, i.e., $b/M<5.19860$ and $b/M>6.23425$. And lensing ring with $3/4<n<5/4$ corresponds to two separated regions with $5.19860<b/M<5.33478$ and $5.36377<b/M<6.23425$. For the photon ring with $n>5/4$, it corresponds to an extremely cramped range of about $5.33478<b/M<5.36377$. Moreover, we also show the case of $c_{13}=0.9$ by the magenta curve in Fig.~\ref{bnf1all}, where the pattern of $n$ keeps the same with that of $c_{13}=0.5$. The big difference is that the peak is shifted toward to large $b/M$ such that $b/M\approx 6.0$. Considering that the peak is corresponded to the inner direct image, one can approach the result that its size increases with the coupling constant $c_{13}$. Furthermore, employing the numerical result, we have the ranges $b/M<5.92667$ and $b/M>6.65141$ for the direct image, $5.92667<b/M<6.00049$ and $6.01088<b/M<6.65141$ for the lensing ring, and $6.00049<b/M<6.01088$ for the photon ring, respectively.

\begin{figure*}
\begin{center}
\subfigure[]{\label{nullf11}
\includegraphics[width=7.5cm]{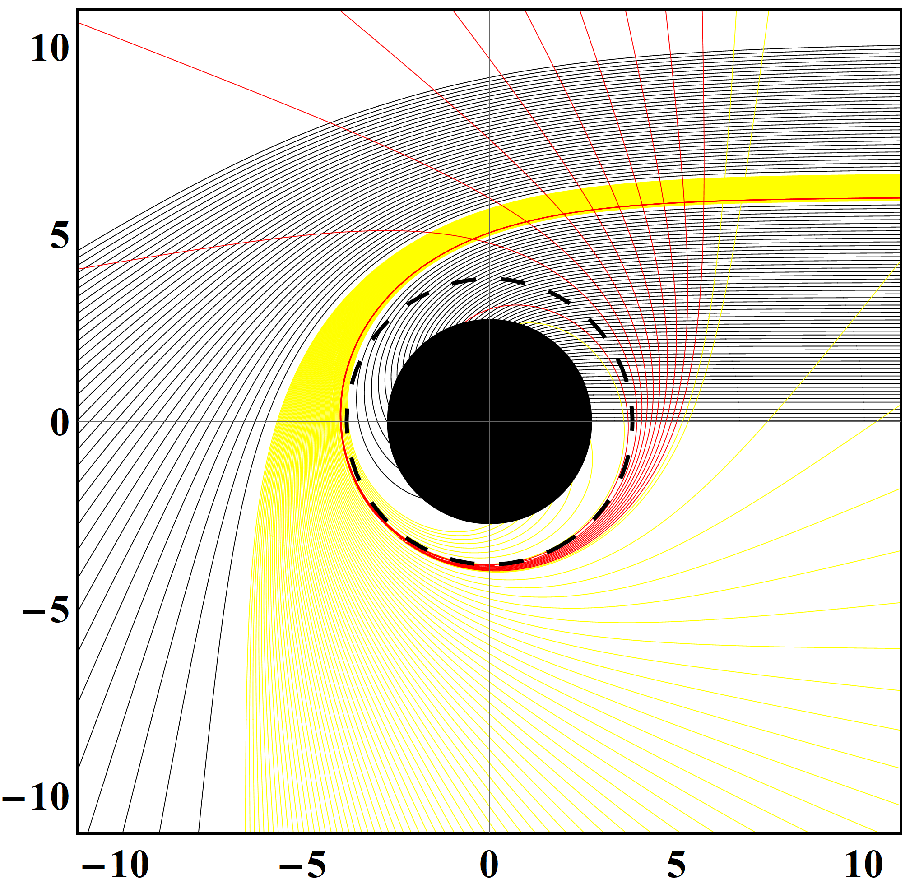}}
\quad
\subfigure[]{\label{nullf21}
\includegraphics[width=7.5cm]{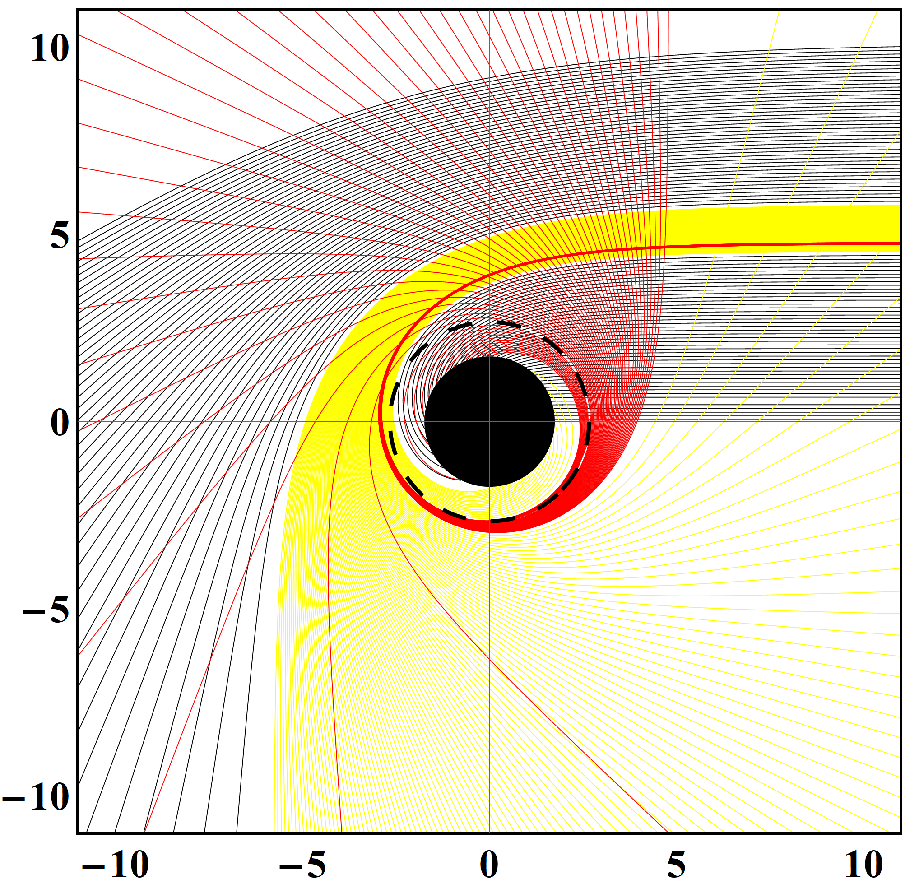}}
\quad
\subfigure[]{\label{nullf22}
\includegraphics[width=7.5cm]{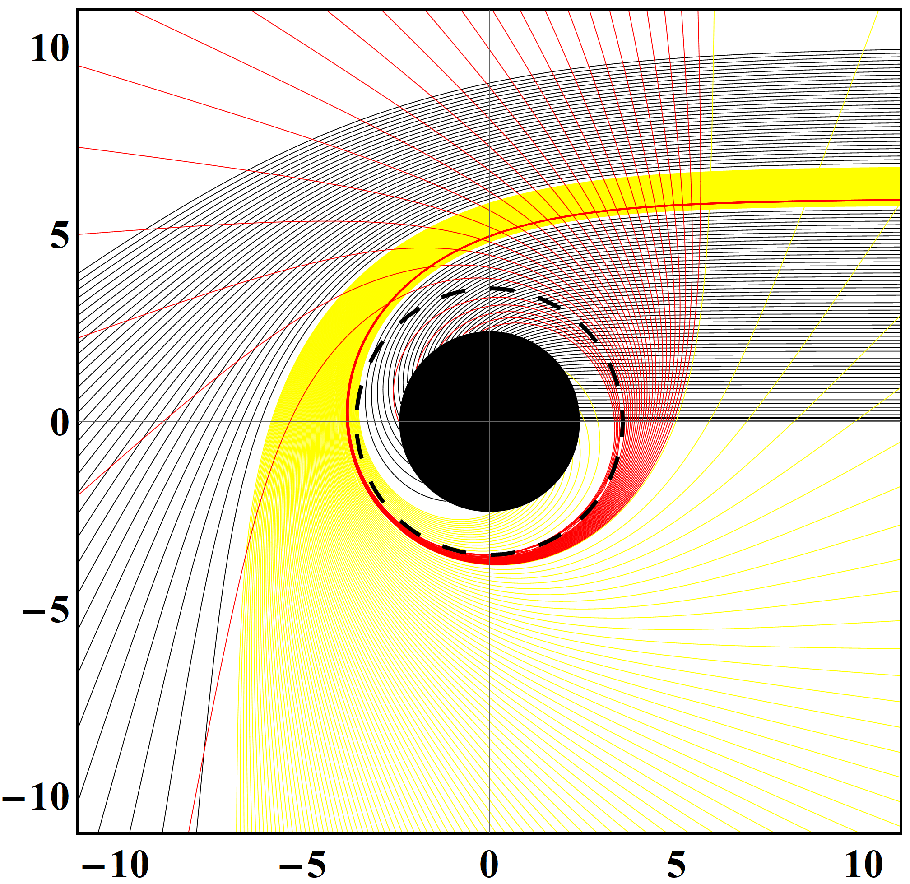}}
\caption{Trajectory of photons around Einstein-\AE ther black hole in the Euclidean polar coordinates ($r,\phi$). The black disks in the center represent the black holes, and the surrounding black dashed circles indicate the photon spheres. Black, yellow, and red curves show the direct image, lensing ring, and photon ring, respectively. (a) For the first Einstein-\AE{}ther black hole solution with $c_{123}\neq 0$, $c_{14}=0$, and $c_{13}=0.9$. (b) For the second Einstein-\AE{}ther black hole solution with $c_{123}=0$, $c_{13}=0$, and $c_{14}=0.9$. (c) For the second Einstein-\AE{}ther black hole solution with $c_{123}=0$, $c_{13}=0.5$, and $c_{14}=0$.}\label{fig2}
\end{center}
\end{figure*}

In addition, we delineate the trajectories of photons with different impact parameters $b/M$ in the Euclidean polar coordinates ($r,\phi$) in Fig.~\ref{nullf11}. Therein, we use the vertical black line to denote the equatorial plane, and assume that the observer is located at the north pole, which is on the very far right of the image. The black disk in the center represents the black hole, and the surrounding black dashed circle indicates the photon sphere. Setting the spacing in the impact parameter as 1/10, 1/100, 1/1000 in the direct image, lensing ring and photon ring bands, respectively, we investigate these trajectories around a Einstein-\AE ther black hole with $c_{13}=0.9$ via ray tracing. From the figure, we can see that for the observer, the direct image described by the black curves, has a large proportion and is divided into two parts. These trajectories in the upper part possess large $b/M$ and intersect the equatorial plane only once. It is obvious that they satisfy $1/4<n<3/4$. For the lower part, it can also be divided into two sub-parts. These trajectories with larger $b/M$ still intersect the equatorial plane once, while the others with smaller $b/M$ has $n<1/4$ and will end at the black hole horizon before crossing the equatorial plane. These trajectories described by the yellow curves give a middle region on the direct image, which stand for the lensing ring occupying only a relatively narrow area. They intersect the equatorial plane twice. And the larger the impact parameter, the farther the distance between these two intersection points. One can also find that, similar to the direct image, the lensing ring is also divided into two parts by the photon ring described by the red curves. Quite differently, these trajectories of the photon ring can intersect the equatorial plane more than three times. Meanwhile, they keep very close with each other and merge into the photon sphere. Another particular feature is that such photon ring is an extremely narrow ring for the observer located at the north pole. This result can also be found in Fig.~\ref{bnf1all}. It is worth to point out that, more details can be revealed by combining with Figs.~\ref{bnf1all} and \ref{nullf11}.

Now let us turn to the trajectory of the photon for the second Einstein-\AE{}ther black hole solution with $c_{123}=0$. After a simple algebraic calculation, the orbit equation reads
\begin{equation}
  \Phi_2(u)=\sqrt{\frac{1}{b^2}+u^2(-1+2Mu)+\frac{M^2u^4(-2c_{13}+c_{14})}{2(-1+c_{13})}}\,.\label{eq:phiu2}
\end{equation}
Taking advantage of Eqs.~\eqref{eq:phi1} and~\eqref{eq:phi2}, one is allowed to calculate the deflection angle for different impact parameters. Further, the total number $n$ of the light trajectory will be obtained.

\begin{table}[h]
  \begin{center}
  \begin{tabular}{|c|c|c|c|c|}
    \hline
    $~c_{13}~$ & $~c_{14}~$  & $n<3/4$  & $3/4<n<5/4$ & $n>5/4$ \\ \hline
    0 & 0.9 & $b<4.55215, ~b>5.80951$ & $4.55215<b<4.75216, ~ 4.80535<b<5.80951$ & $4.75216<b<4.80535$  \\
    0 & 1.8 & $b<3.84574, ~b>5.39259$  & $3.84574<b<4.14864,~ 4.25082<b<5.39259$ & $4.14864<b<4.25082$  \\
    \hline
    0.5 & 0 & $b<5.78364, ~b>6.83755$ & $5.78364<b<5.93354, ~ 5.96292<b<6.83755$  & $5.93354<b<5.96292$ \\
    0.9 & 0 & ~$b<9.15824,~ b>10.16178$ ~& ~$9.15824<b<9.29209,~ 9.31045<b<10.16178$ ~& ~$9.29209<b<9.31045$~ \\  \hline
  \end{tabular}
  \caption{Ranges of the impact parameter $b$ for the direct image, lensing ring, and photon ring, respectively. We choose the second Einstein-\AE{}ther black hole solution with different values of the coupling constants.}\label{tablef2}
  \end{center}
\end{table}
For this case, we should note that the coupling constant $c_{14}$ will significantly contribute to the photon behavior. In order to clearly show it, we consider two specific cases in Fig.~\ref{bnf21all} with the coupling constants $c_{13}=0$ while $c_{14}$ is set to $0.9$ (blue dot dashed curve) and $1.8$ (red solid curve), respectively. From Fig.~\ref{bnf21all}, we observer that these two curves share the similar pattern. With the increase of $b/M$, the number $n$ increases slowly at first, and then gets a rapid increase until the critical impact parameter $b_{c}/M$ is approached. Finally, $n$ decreases and tends to $n=1/2$. Although the behavior of $n$ does not change for different values of $c_{14}$, the peak is shifted to the left by it. In Table~\ref{tablef2}, we list the ranges of $b/M$ of the direct image, lensing ring, and photon ring for different values of the coupling constants $c_{13}$ and $c_{14}$. From numerical results, we observe that $c_{14}$ does have a distinct effect on these three images. For example, when $c_{14}=0.9$, the radius of the internal direct image is 4.55215, and the widths of the lensing ring and photon ring are 1.25736 and 0.05319, respectively. While for the case $c_{14}=1.8$, they are 3.84574, 1.54685, and 0.10218, respectively. It indicates that the coupling constant $c_{14}$ narrows the internal direct image, while widens the lensing ring and photon ring.

Apart from $c_{14}$, the coupling constant $c_{13}$ also appears in the second Einstein-\AE{}ther black hole solution, and which will also affect the motion of the photon. When $c_{14}=0$, we plot the number $n$ in Fig.~\ref{bnf22all} for $c_{13}=0.5$ (purple dot dashed curve) and 0.9 (magenta solid curve), respectively. The behaviors are also similar to that shown in Figs.~\ref{bnf1all} and \ref{bnf21all}. In particular, with the increase of $c_{13}$, the peak is shifted to the right. This implies that the size of the internal direct image increases with $c_{13}$. From another point of view, the photons are more harder to escape from the black hole for a large value of $c_{13}$. For instance, the photon with $b/M=8$ can escape from the black hole for $c_{13}=0.5$, while cannot for $c_{13}=0.9$.

The photon trajectories are shown in Fig.~\ref{nullf21} for $c_{123}=0$, $c_{13}=0$, and $c_{14}=0.9$ and Fig.~\ref{nullf22} for $c_{123}=0$, $c_{13}=0.5$, and $c_{14}=0$. Both them share a similar pattern as that for the first Einstein-\AE ther black hole solution given in Fig. \ref{nullf11}. For each case, the direct image still takes up most of the space, and it is divided into two parts by the lensing ring. Within this ring, these photons with larger impact parameter $b/M$ are visible, while those with smaller $b/M$ are invisible to the observer. The lensing ring is composed of photons with $3/4<n<5/4$, which is separated into two parts by the photon ring as well. The part far away from the black hole is wider while the other is narrower. The photon ring has a tiny width. If one goes back against the observer's line of sight, these light rays will asymptotically approach the photon sphere.

\begin{table}[h]
  \begin{center}
  \begin{tabular}{|c|c|c|c|c|c|c|c|}
    \hline
    $~$ & Coupling constants & $r_0$ & $r_p$ &  $b_c$  & ~$w_{di\text{(internal)}}$ ~ & $w_{lr}$  & $w_{pr}$     \\ \hline
    GR &  $c_{1}=c_{2}=c_{3}=c_{4}=0$ &   2   &   3   & 5.19615 & 5.01514   & 1.15243   & 0.04013  \\ \hline
    Case I &~ $c_{123}\neq 0$, $c_{13}=0.9$, $c_{14}=0$ ~& ~2.73906~ & ~3.81840~ & ~6.00169~ & 5.92667 & ~0.72474~ & ~0.01039~ \\ \hline
    Case II &$c_{123}=0$, $c_{13}=0$, $c_{14}=0.9$& 1.74162 & 2.66190 & 4.76431 & 4.55215 & 1.25736 & 0.05320 \\ \hline
   Case III & $c_{123}=0$, $c_{13}=0.5$, $c_{14}=0$& 2.42421 & 3.56155 & 5.93912 & 5.78364 & 1.05391 & 0.02938 \\  \hline
  \end{tabular}
  \caption{The values of the radii of event horizon $r_0$ and photon sphere $r_p$, the critical impact parameter $b_c$, and the widths of the direct image (internal) $w_{di\text{(internal)}}$, lensing ring $w_{lr}$, and photon ring $w_{pr}$ for GR and Einstein-\AE ther theory with different coupling constants.}\label{tableGE}
  \end{center}
\end{table}
Before ending this section, we summarize the effects of these coupling constants on the motion of the photon, and then discuss the differences in the behavior of the photon between Einstein-\AE ther theory and GR. For the first black hole solution (\ref{eq:f1}) in Einstein-\AE ther theory with $c_{123}\neq 0$ and $c_{14}=0$, only the coupling constant $c_{13}$ is nonzero. The results show that the radii of the event horizon $r_{0}$, photon sphere $r_p$, and the critical impact parameter $b_c$ increase with $c_{13}$. As a result, a larger internal dark area and a thinner lensing ring will be produced by large value of $c_{13}$.

For the second Einstein-\AE{}ther black hole solution (\ref{eq:f2}) with $c_{123}=0$, there are two coupling constants $c_{13}$ and $c_{14}$. When $c_{13}$ is fixed, $r_0$, $r_p$, and $b_c$ decrease with $c_{14}$. Moreover, the corresponding radius of the internal dark area also decreases, while the width of the lensing ring increases with it. If one fixes $c_{14}$ instead, the coupling constant $c_{13}$ plays a reverse effect on these quantities. Finally, in order to compare with the GR black hole, we list the values of $r_0$, $r_p$, $b_c$, and the widths of the internal direct image $w_{di\text{(internal)}}$, lensing ring $w_{lr}$, and photon ring $w_{pr}$ in Table~\ref{tableGE} for different coupling constants. The results show that photons behave quite differently in Einstein-\AE ther theory and GR. For the case I and case III, $r_0$, $r_p$, $b_c$, and $w_{di\text{(internal)}}$ in Einstein-\AE ther theory are larger than that in GR, but $w_{lr}$ and $w_{pr}$ are smaller. On the other hand, for the case II, we will arrive at a converse conclusion. The black hole in GR has larger event horizon, photon sphere, and the dark area corresponding to the direct image, while narrower lensing ring and photon ring.

\section{Image of black hole surrounded by a thin disk emission}
\label{image}

The general discussion of black hole shadow is based on the assumption that photons are all from infinity and deflected by gravity when they pass through the black hole. Photons falling into the black hole will not reach the observer and thus form a shadow. Its boundary is determined by the critical impact parameter. In fact, there will be a massive amount of accretion material around black holes in our universe. Therefore we will consider a specific case that there is an optically thin and geometrically thin accretion disk on the equatorial plane of the black hole, from which photons are emitted. Under this assumption, the effects of absorption and reflection of the photons can be ignored. Besides, we will further assume that there is a static observer located at the north pole of the black hole.

In this section, three simple toy-model functions will be applied to study the specific intensity of emission $I_{\text{em}}$, which is only related to the radial coordinate $r$. In the first model, we assume that the emission suddenly increases, peaks, and then decays sharply at the innermost stable circular orbit (ISCO), while with no emission inside it. The function of the intensity of emission reads
\begin{equation}\label{eq:iem1}
  I_{\text{em}1}(r)=\Bigg\{
   \begin{array}{lcl}
      I_0\left( \frac{1}{r-(r_{\text{ISCO}}-1)} \right)^2,&~&r>r_{\text{ISCO}},\\
      0,&~&r\leq r_{\text{ISCO}},
   \end{array}
\end{equation}
where the radius $r_{\text{ISCO}}$ of ISCO is greater than that of the photon sphere $r_p$. In the second model, the emission is assumed to appear suddenly at the photon sphere, then decay suppressed by the third power
\begin{equation}\label{eq:iem2}
  I_{\text{em}2}(r)=\Bigg\{
   \begin{array}{lcl}
      I_0 \left( \frac{1}{r-(r_{p}-1)} \right)^3,&~&r>r_{p},\\
      0,&~&r\leq r_{p}.
   \end{array}
\end{equation}
The third model considers the emission starting from the event horizon and then following by a relatively slowly decay towards the ISCO in contrast to the first two models. Its intensity of emission can be expressed as
\begin{equation}\label{eq:iem3}
  I_{\text{em}3}(r)=\Bigg\{
   \begin{array}{lcl}
      I_0 \frac{\frac{\pi}{2}-\arctan(r-(r_{\text{ISCO}}-1))}{\frac{\pi}{2}-\arctan(r_0-(r_{\text{ISCO}}-1))},&~&r>r_0,\\
      0,&~&r\leq r_0.
   \end{array}
\end{equation}
It is thought that the image of the black hole viewed by the observer closely depends on these emission models. Now, we first discuss the relationship between the observed specific intensity $I_{\text{obs}}$ and the emitted specific intensity $I_{\text{em}}$. Along the light ray, the quantity $I_v/{v^3}$ is conserved, thus we have
\begin{equation}
  \frac{I_{\text{em},v}}{v^3}=\frac{I_{\text{obs},v'}}{v'^3},
\end{equation}
where $v$ and $v'$ denote the local frequencies of the photons measured at the locations of the emitted source and the observer, respectively. Due to the gravitational redshift, one has $v'/v=\sqrt{f(r)}$ . Therefore, the observed specific intensity can be expressed as
\begin{equation}
  I_{\text{obs},v'}=f^{3/2}(r)I_{\text{em},v}.
\end{equation}
Further, we obtain the total observed specific intensity by integrating the above equation with respect to the frequency:
\begin{equation}
  I_{\text{obs}}=\int I_{\text{obs},v'}\text{d}v'=\int f^2 (r) I_{\text{em},v} \text{d}v=f^2 (r) I_{\text{em}}(r).\label{eq:iobsr}
\end{equation}

\begin{figure*}
\begin{center}
\subfigure[]{\label{brf11}
\includegraphics[width=5.5cm]{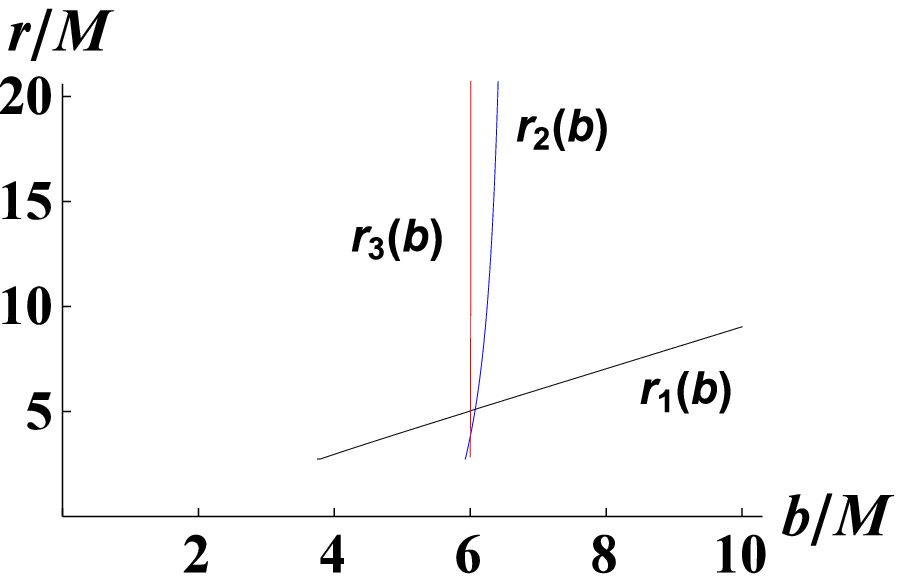}}
\quad
\subfigure[]{\label{brf21}
\includegraphics[width=5.5cm]{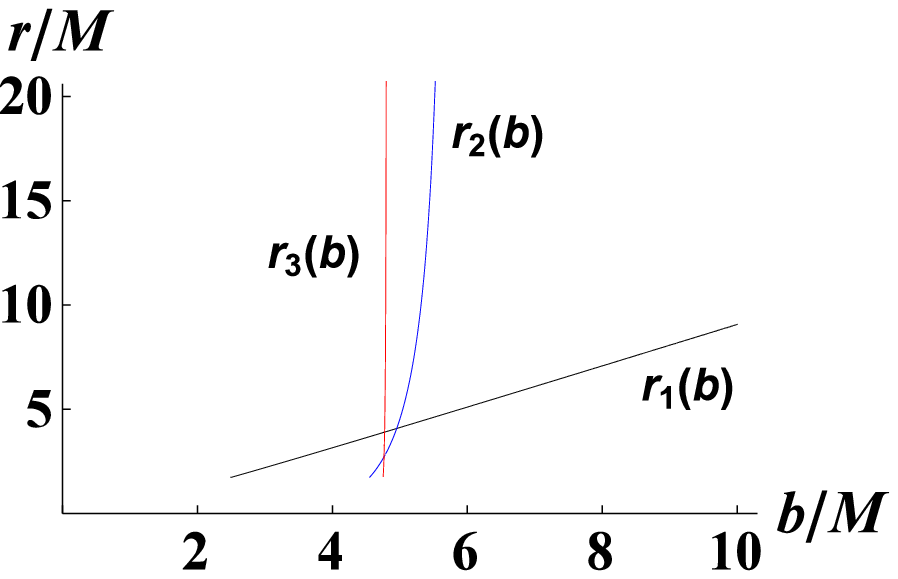}}
\quad
\subfigure[]{\label{brf22}
\includegraphics[width=5.5cm]{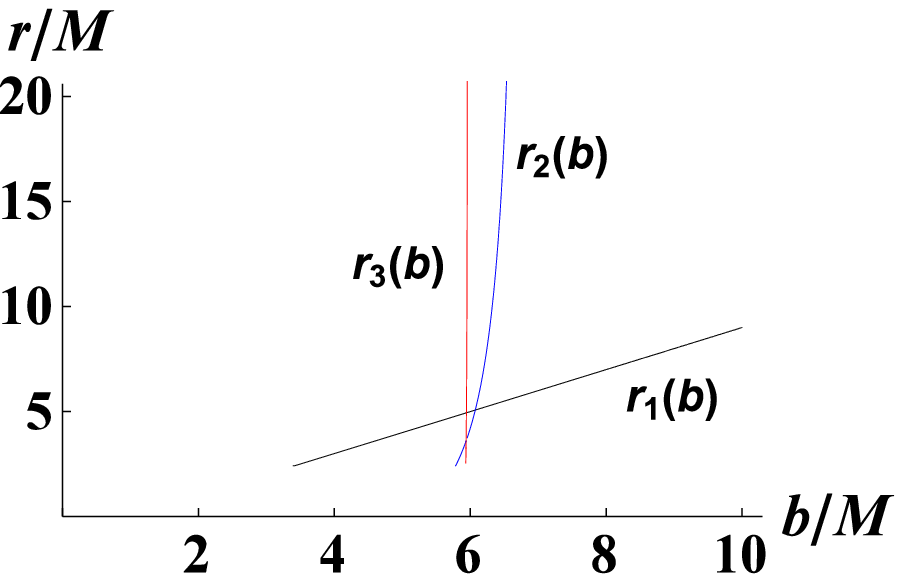}}
\caption{The first three transfer functions in Einstein-\AE ther spacetime. The black, blue, and red curves represent the radial coordinate of the first, second, and third intersections with the disk. (a) For the first Einstein-\AE{}ther black hole solution with $c_{123}\neq 0$, $c_{14}=0$, and $c_{13}=0.9$. (b) For the second Einstein-\AE{}ther black hole solution with $c_{123}=0$, $c_{13}=0$, and $c_{14}=0.9$. (c) For the second Einstein-\AE{}ther black hole solution with $c_{123}=0$, $c_{13}=0.5$, and $c_{14}=0$.}\label{fig3}
\end{center}
\end{figure*}

So far, we have given the emitted specific intensities through three simple models \eqref{eq:iem1}, \eqref{eq:iem2}, and \eqref{eq:iem3}, as well as the relationship between the observed specific intensity and the emitted one \eqref{eq:iobsr}. It is also important to note that the photon will get extra brightness from the intersection as passing through the disk. For photons with $1/4<n<3/4$, they intersect the disk once from the front side. When $3/4<n<5/4$, the photon trajectory will be bent around the black hole, and photons will intersect the disk from the back side, gaining a second extra brightness. When $n>5/4$, photons will intersect the disk at least three times. As a result, the larger $n$ is, the more extra brightness photons will obtain from the disk. Obviously, the observed specific intensity finally received by the observer is the sum of the brightness obtained from all intersections:
\begin{equation}
  I_{\text{obs}}(b)=\sum_{m} f^2 (r) I_{\text{em}}|_{r=r_m(b)}.\label{eq:iobsb}
\end{equation}
Note that $r_m(b)$ is the transfer function denoting the radial position of the $m$-th intersection with the disk for a given impact parameter $b$. The slope of the transfer function $\text{d}r/\text{d}b$ is called the demagnification factor, which characterizes the demagnified scale of the transfer function. Note that the transfer function is independent of the specific emitted models. We plot the first three transfer functions in Fig.~\ref{fig3} for different values of the coupling constants.

For the first Einstein-\AE{}ther black hole solution with $c_{13}=0.9$, we exhibit the transfer functions in Fig.~\ref{brf11}. The first transfer function described by the black curve corresponds to the direct image of the disk. The slope is close to 1 indicating that the image profile is just a redshifted source profile. The second transfer function represented by the blue curve corresponds to the lensing ring. The very large slope indicates a high demagnification of image. Hence, the total flux contributed by the lensing ring is small. The red curve with an almost infinite slope represents the third transfer function, corresponding to an extremely demagnified image, the photon ring. It contributes less to the total flux. Moreover, it is not difficult to find that the first transfer function starts at $b/M=3.8$, and the lensing ring and photon ring appear near $b/M=6$. Figs.~\ref{brf21} and \ref{brf22} are for the second Einstein-\AE{}ther black hole solution, respectively, has $c_{13}=0, ~ c_{14}=0.9$ and $c_{13}=0.5,~ c_{14}=0$. We observe that the values of these coupling constants mainly influence the starting point of these transfer functions. A universal result is that the slopes of the second and third transfer functions are very large, leading to a small contribution to the total flux.

\begin{figure*}
\begin{center}
\subfigure[]{\label{iemf11}
\includegraphics[width=5.5cm]{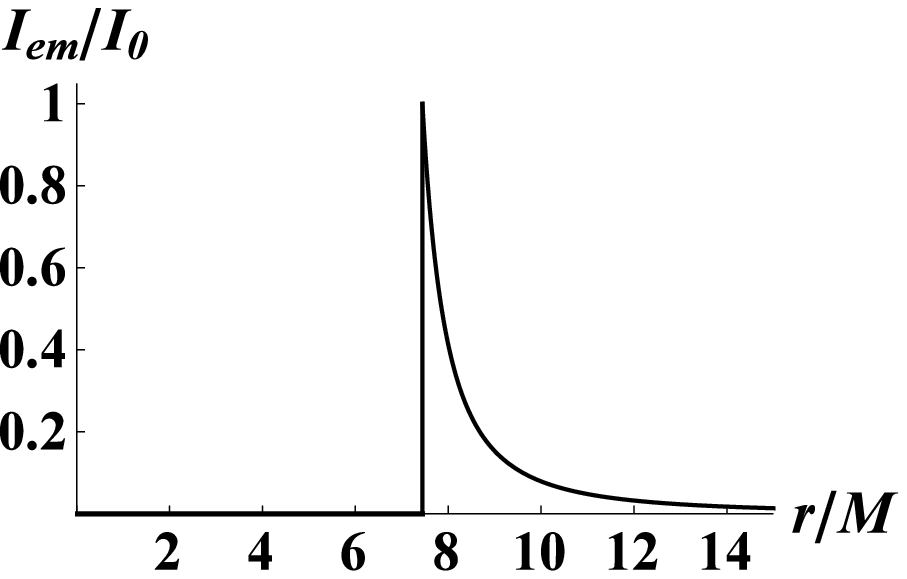}}
\quad
\subfigure[]{\label{iemf12}
\includegraphics[width=5.5cm]{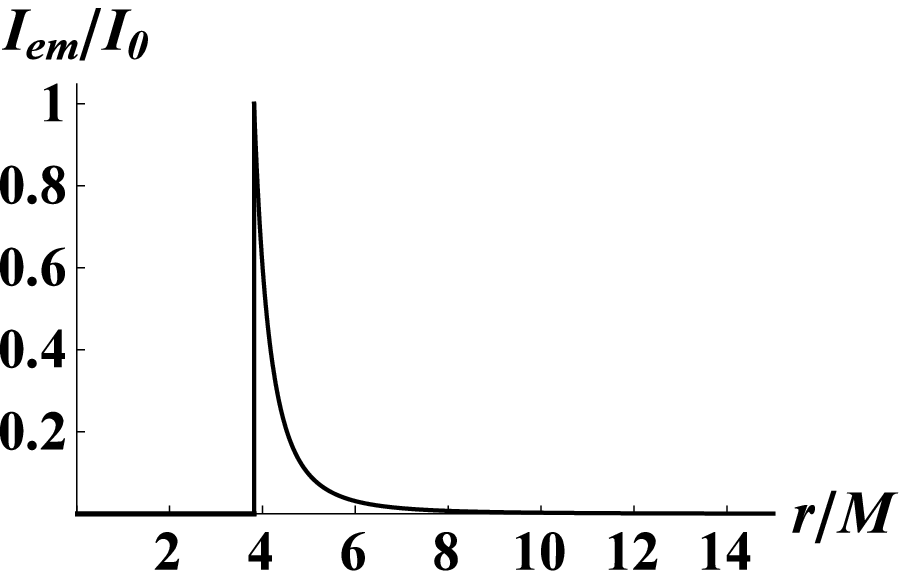}}
\quad
\subfigure[]{\label{iemf13}
\includegraphics[width=5.3cm]{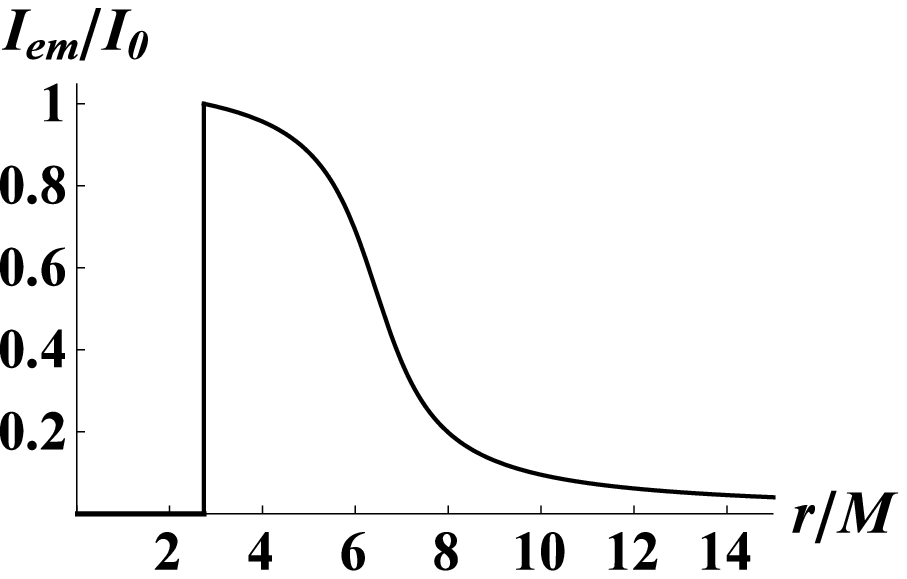}}\\
\subfigure[]{\label{iobf11}
\includegraphics[width=5.5cm]{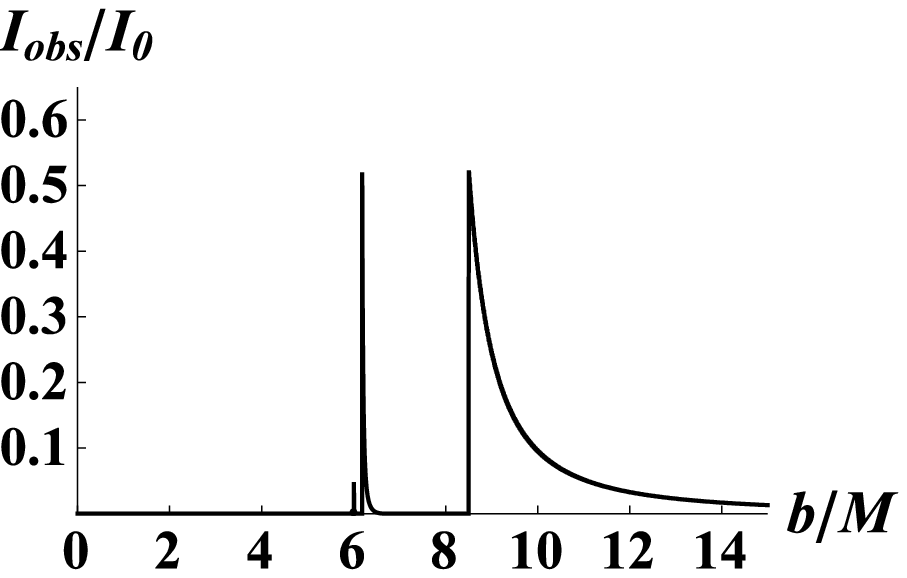}}
\quad
\subfigure[]{\label{iobf12}
\includegraphics[width=5.5cm]{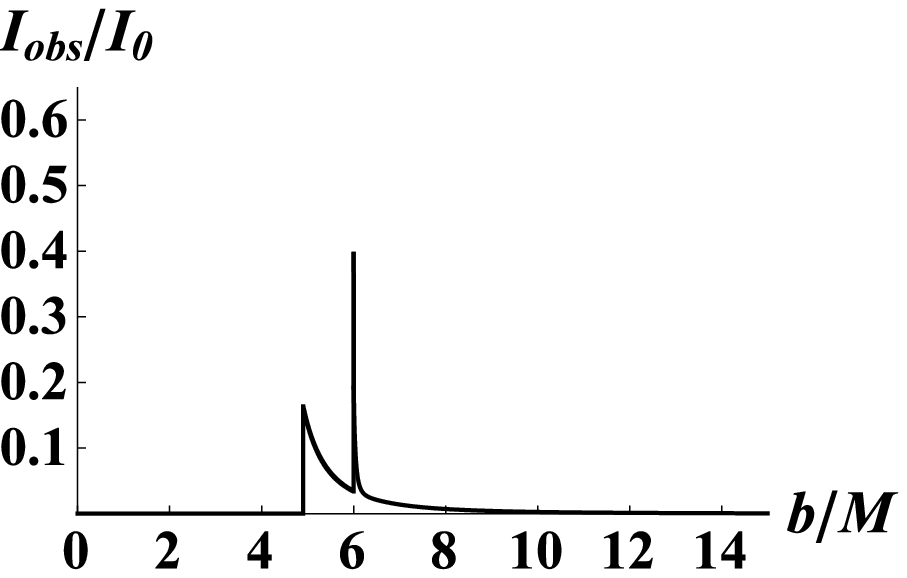}}
\quad
\subfigure[]{\label{iobf13}
\includegraphics[width=5.5cm]{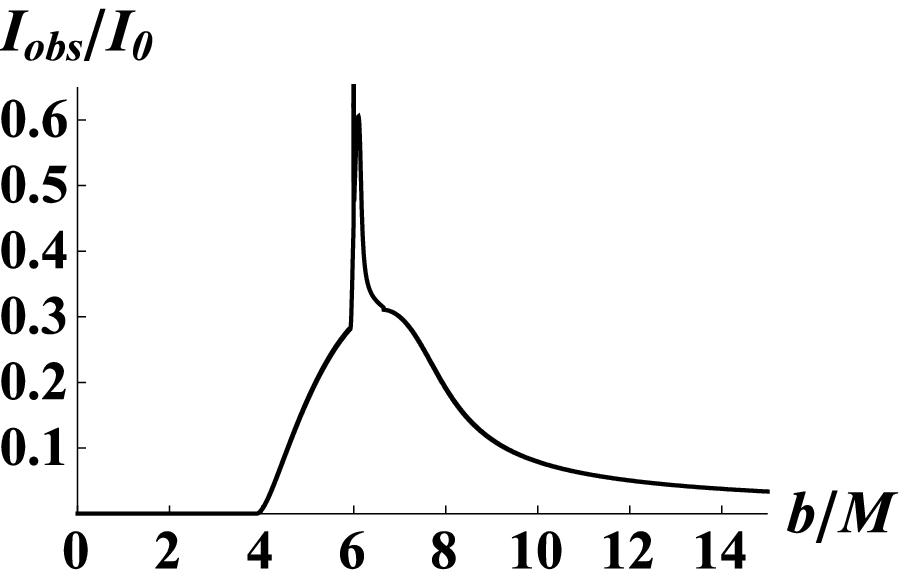}}
\\
\subfigure[]{\label{shadowf11}
\includegraphics[width=5.5cm]{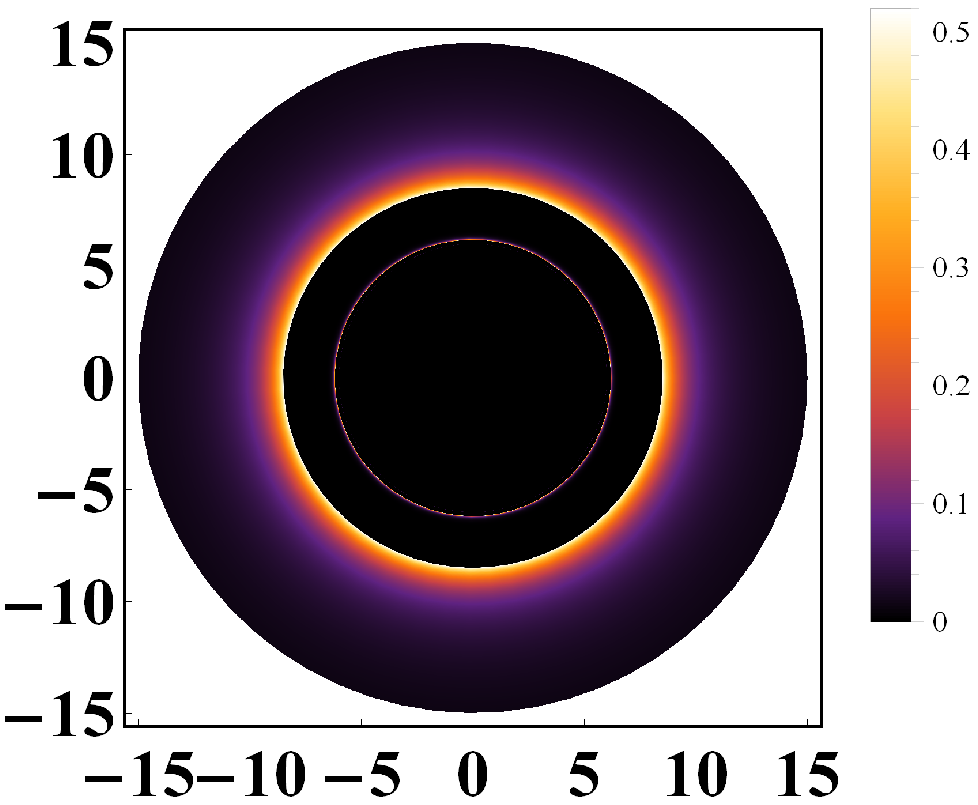}}
\quad
\subfigure[]{\label{shadowf12}
\includegraphics[width=5.5cm]{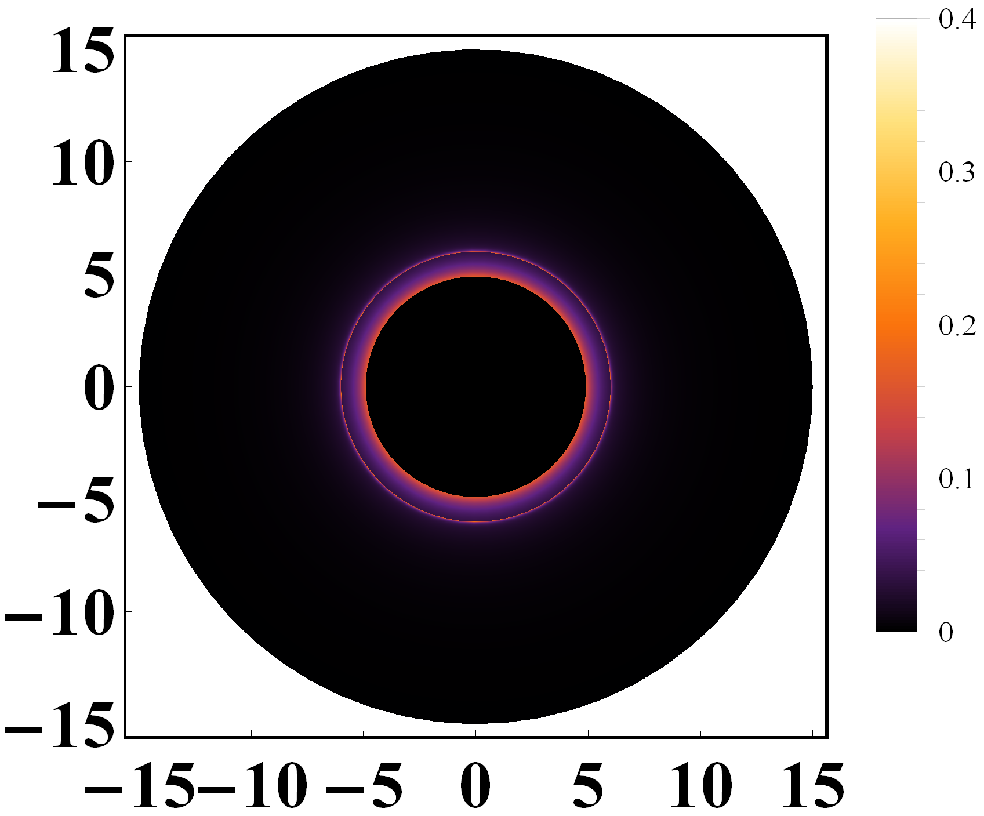}}
\quad
\subfigure[]{\label{shadowf13}
\includegraphics[width=5.5cm]{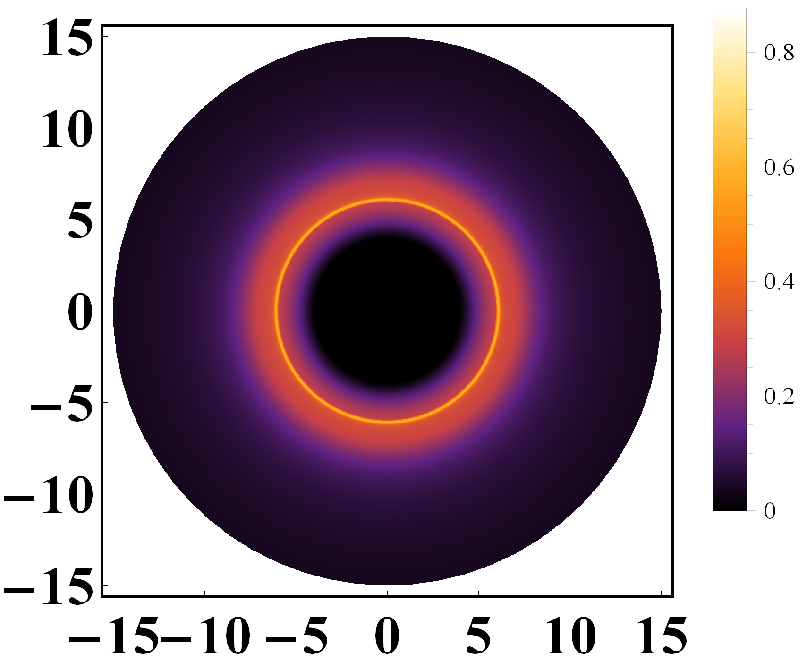}}
\caption{Optical appearances of the thin disk with different emission profiles near the Einstein-\AE ther black hole for the first Einstein-\AE{}ther black hole solution with $c_{123}\neq 0$, $c_{14}=0$, and $c_{13}=0.9$. The top line shows the profiles of three emissions $I_{\text{em}}(r)$. The middle line shows the observed specific intensities $I_{\text{obs}}$ as a function of $b/M$. The bottom line is the density plots of $I_{\text{obs}}$.}\label{shadow1}
\end{center}
\end{figure*}
\begin{figure*}
\begin{center}
\subfigure[]{\label{iemf21}
\includegraphics[width=5.5cm]{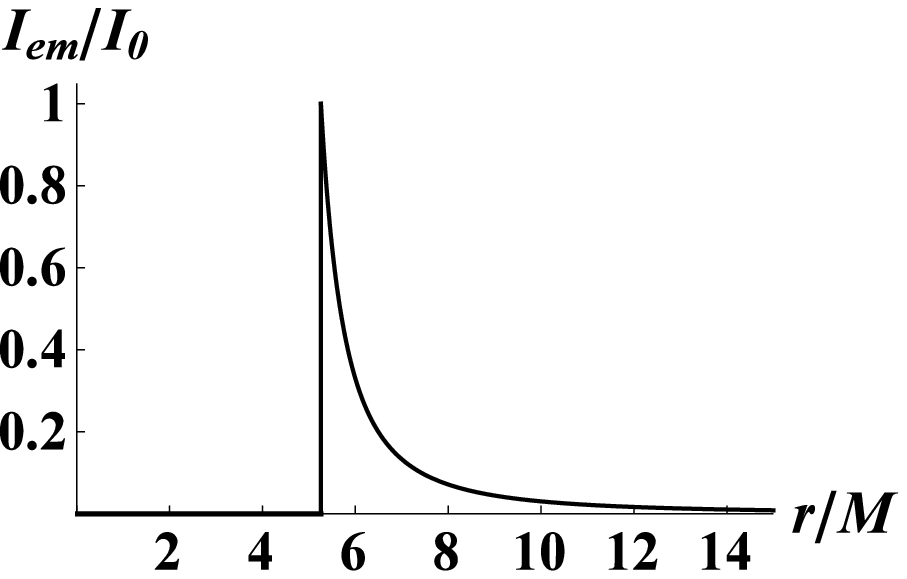}}
\quad
\subfigure[]{\label{iemf22}
\includegraphics[width=5.5cm]{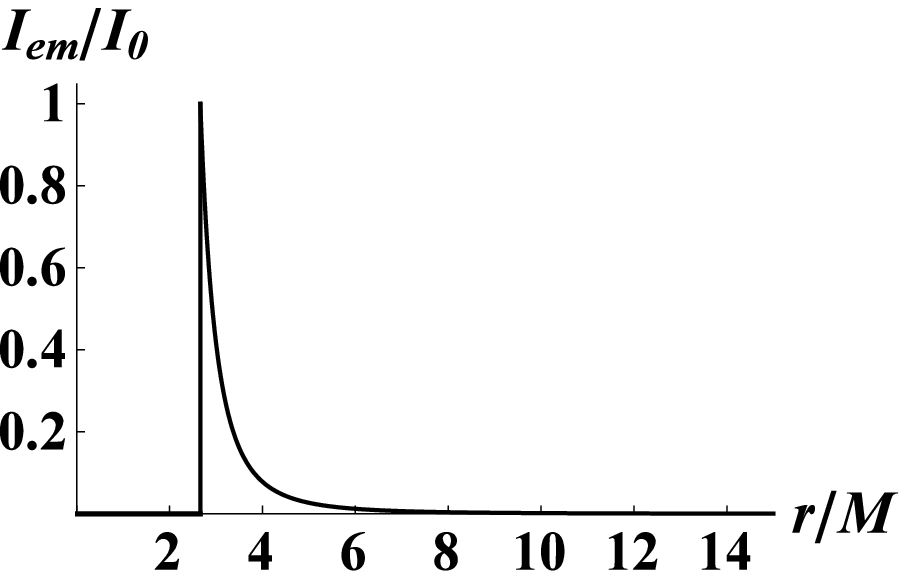}}
\quad
\subfigure[]{\label{iemf23}
\includegraphics[width=5.5cm]{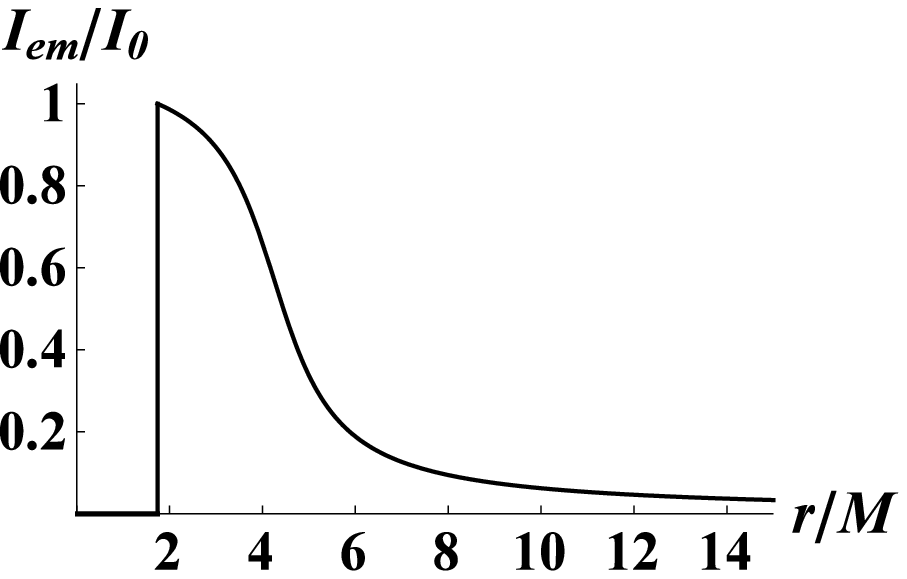}}\\
\subfigure[]{\label{iobf21}
\includegraphics[width=5.5cm]{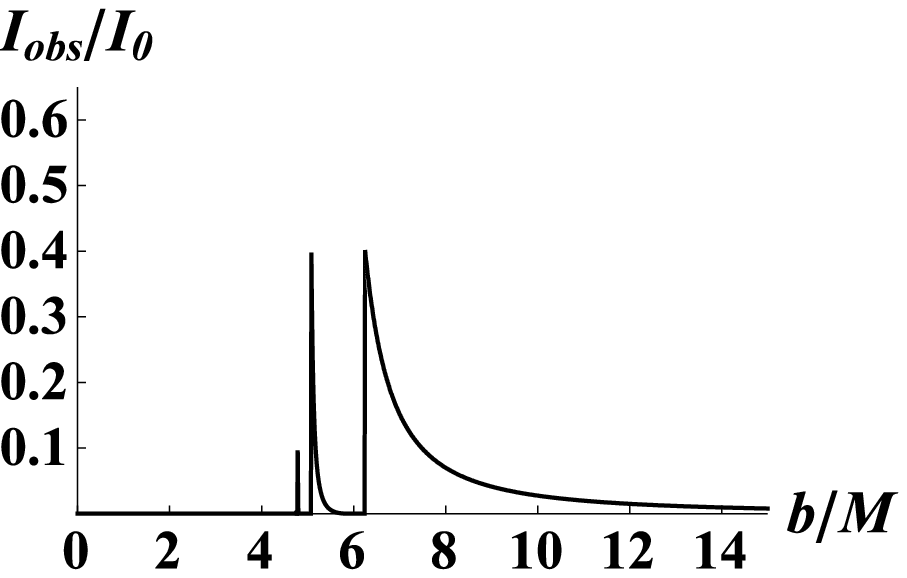}}
\quad
\subfigure[]{\label{iobf22}
\includegraphics[width=5.5cm]{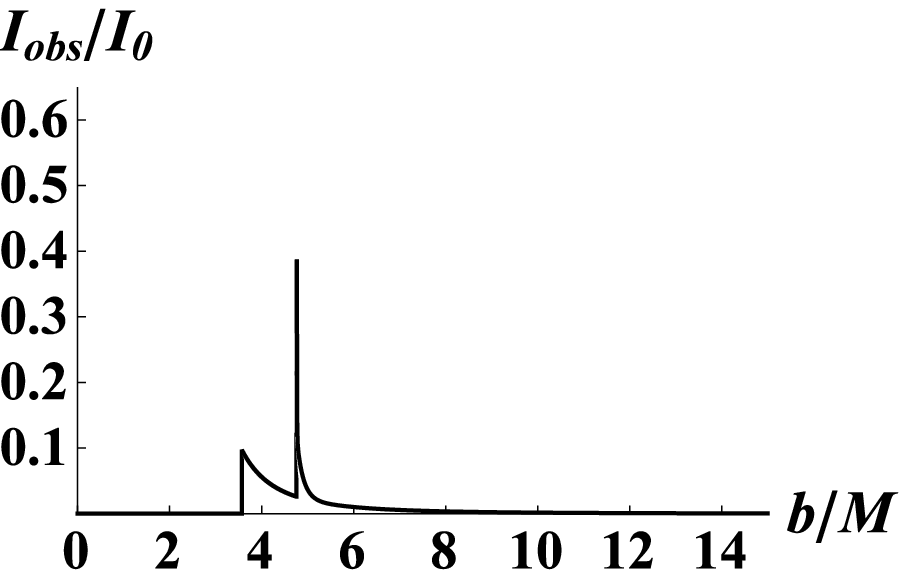}}
\quad
\subfigure[]{\label{iobf23}
\includegraphics[width=5.5cm]{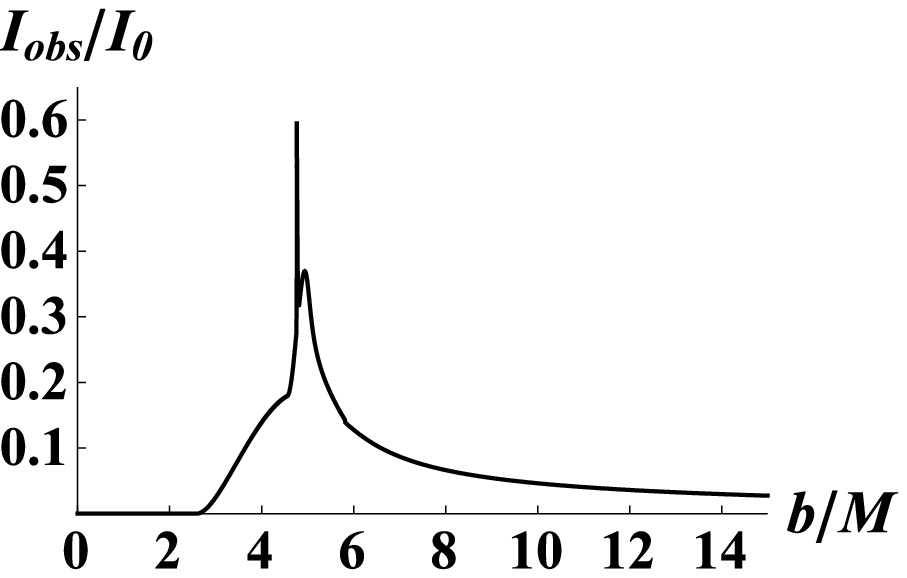}}
\\
\subfigure[]{\label{shadowf21}
\includegraphics[width=5.5cm]{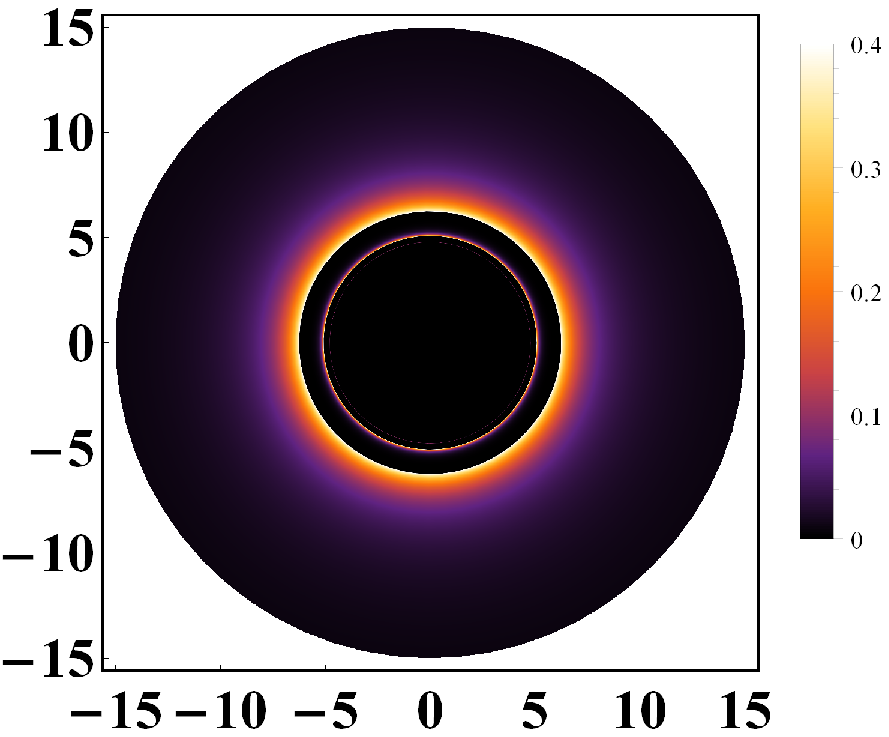}}
\quad
\subfigure[]{\label{shadowf22}
\includegraphics[width=5.5cm]{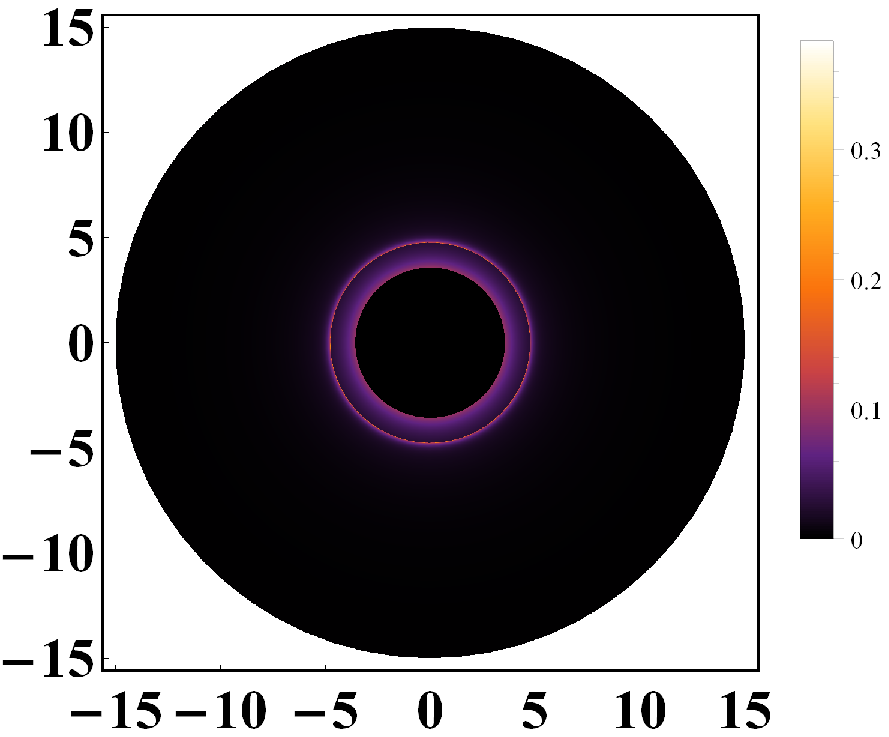}}
\quad
\subfigure[]{\label{shadowf23}
\includegraphics[width=5.5cm]{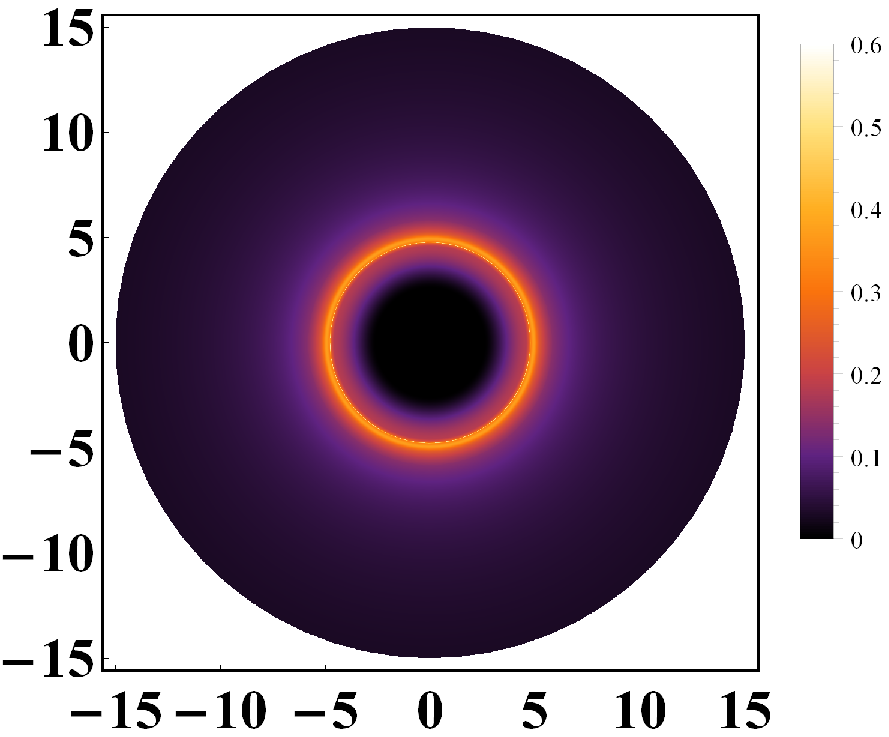}}
\caption{Optical appearances of the thin disk with different emission profiles near the Einstein-\AE ther black hole for the second Einstein-\AE{}ther black hole solution with $c_{123}=0$, $c_{13}=0$, and $c_{14}=0.9$. The top line shows the profiles of three emissions $I_{\text{em}}(r)$. The middle line shows the observed specific intensities $I_{\text{obs}}$ as a function of $b/M$. The bottom line is the density plots of $I_{\text{obs}}$.}\label{shadow2}
\end{center}
\end{figure*}
\begin{figure*}
\begin{center}
\subfigure[]{\label{iemf24}
\includegraphics[width=5.5cm]{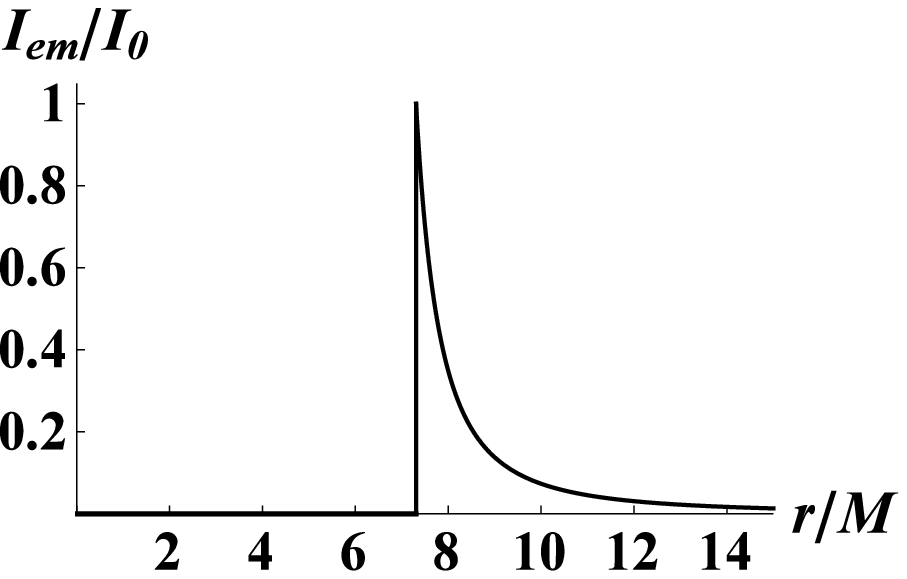}}
\quad
\subfigure[]{\label{iemf25}
\includegraphics[width=5.5cm]{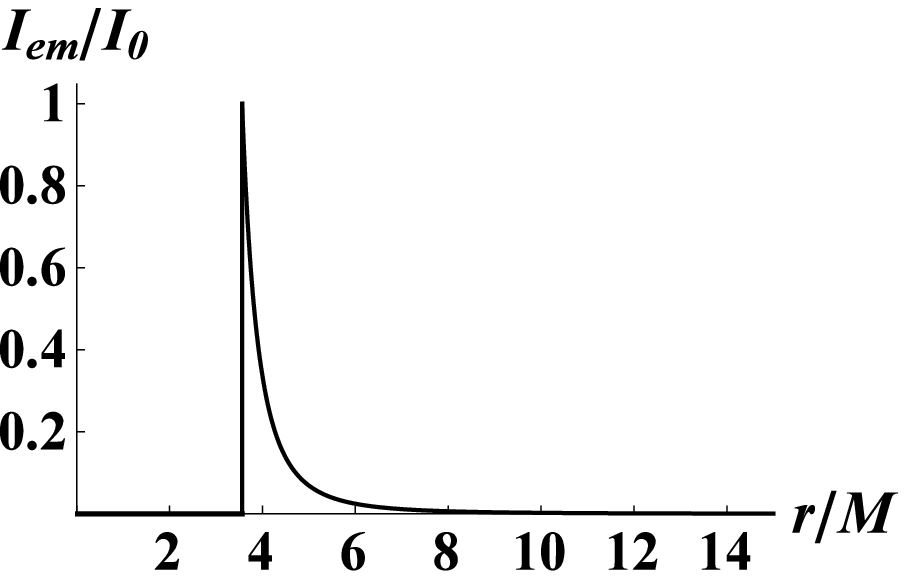}}
\quad
\subfigure[]{\label{iemf26}
\includegraphics[width=5.5cm]{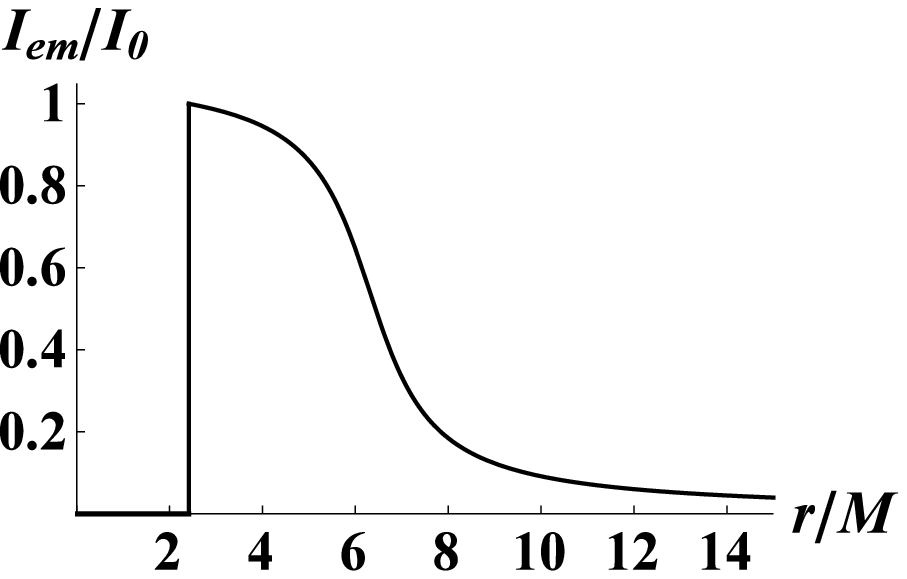}}\\
\subfigure[]{\label{iobf24}
\includegraphics[width=5.5cm]{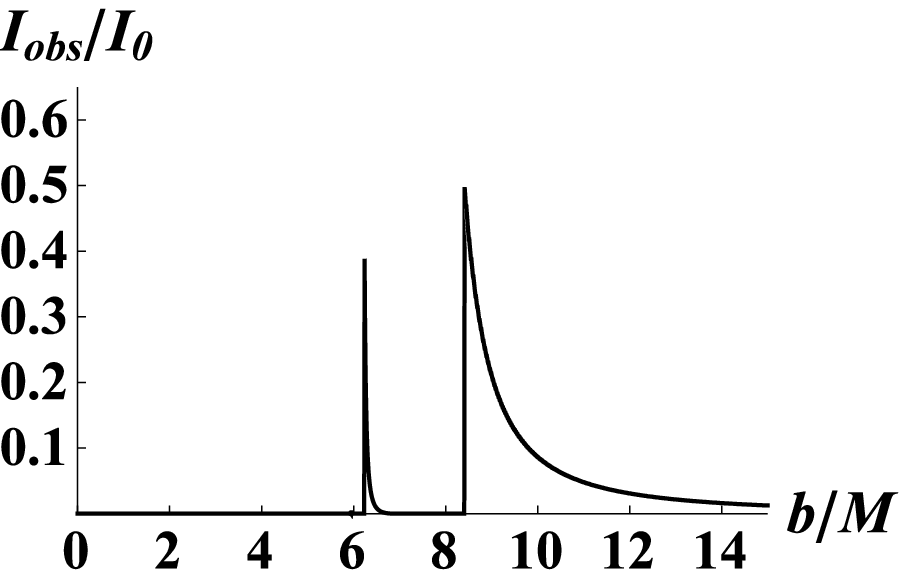}}
\quad
\subfigure[]{\label{iobf25}
\includegraphics[width=5.5cm]{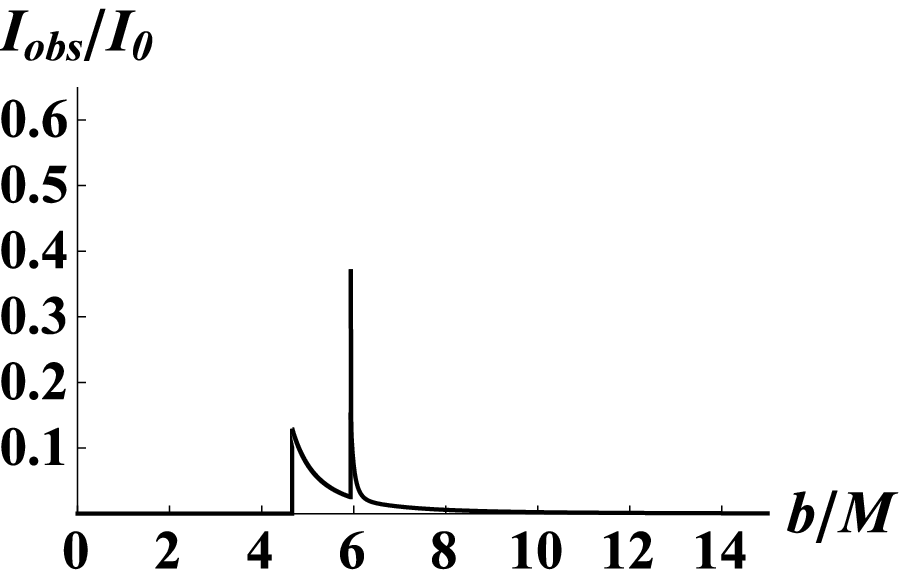}}
\quad
\subfigure[]{\label{iobf26}
\includegraphics[width=5.5cm]{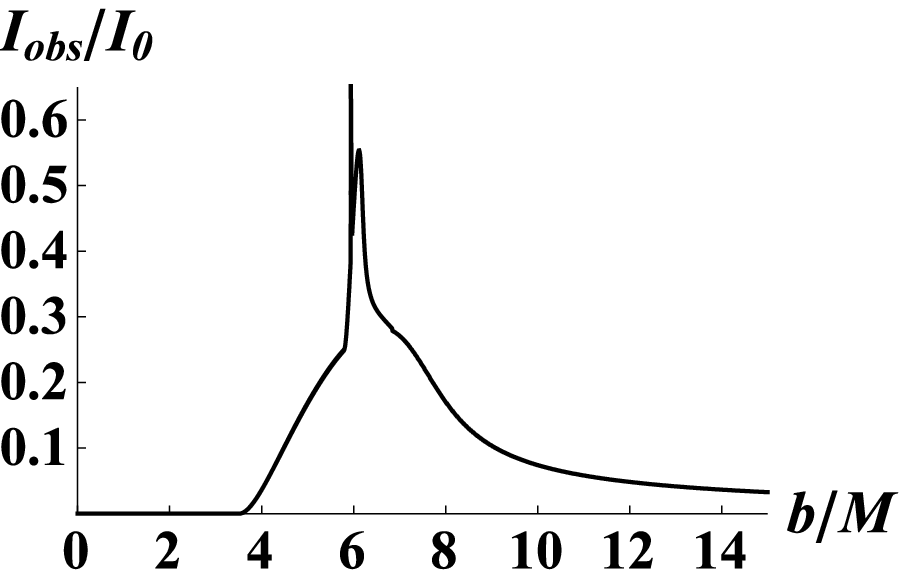}}
\\
\subfigure[]{\label{shadowf24}
\includegraphics[width=5.5cm]{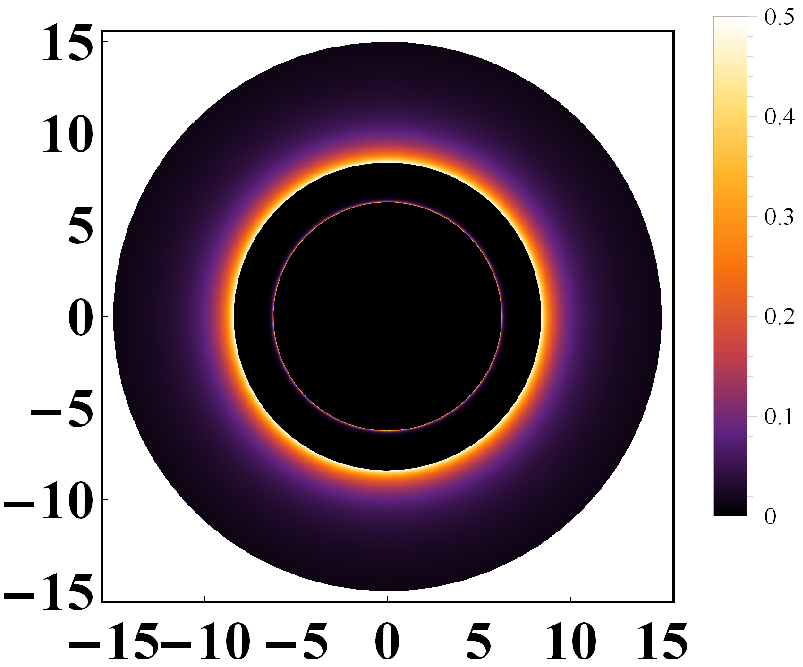}}
\quad
\subfigure[]{\label{shadowf25}
\includegraphics[width=5.5cm]{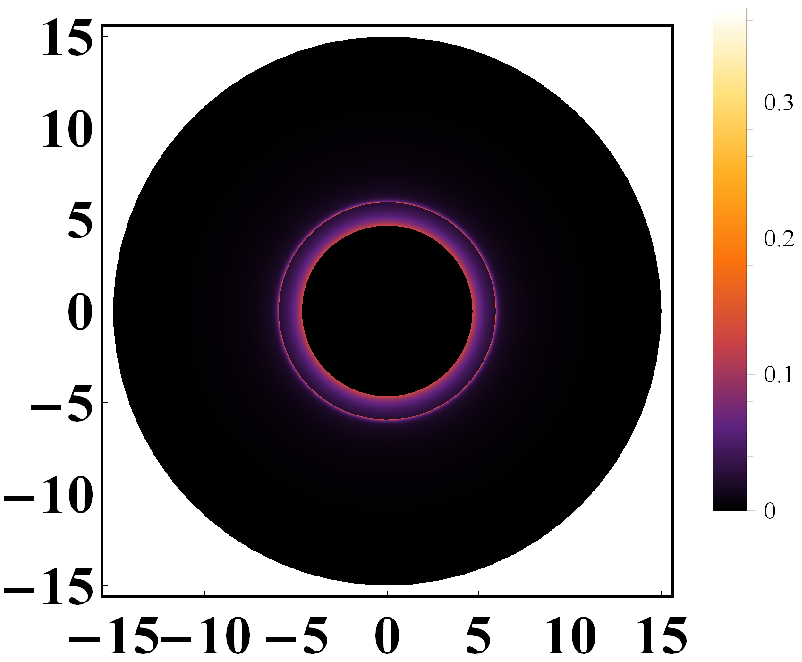}}
\quad
\subfigure[]{\label{shadowf26}
\includegraphics[width=5.5cm]{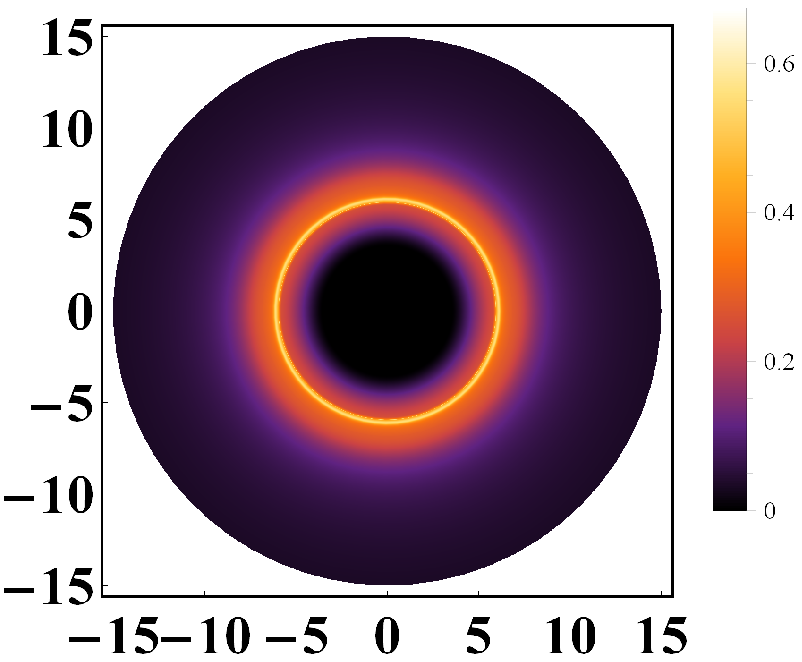}}
\caption{Optical appearances of the thin disk with different emission profiles near the Einstein-\AE ther black hole for the second Einstein-\AE{}ther black hole solution with $c_{123}=0$, $c_{13}=0.5$, and $c_{14}=0$. The top line shows the profiles of three emissions $I_{\text{em}}(r)$. The middle line shows the observed specific intensities $I_{\text{obs}}$ as a function of $b/M$. The bottom line is the density plots of $I_{\text{obs}}$.}\label{shadow3}
\end{center}
\end{figure*}

Basing on these results, we describe the optical appearance of the black hole surrounded by a thin disk in Figs.~\ref{shadow1}, \ref{shadow2}, and \ref{shadow3}. In the follows, we shall give a detailed discussion on the effects of these coupling constants.

For the first Einstein-\AE{}ther black hole solution with $c_{123}\neq 0$, $c_{14}=0$, and $c_{13}=0.9$, the results are presented in Fig.~\ref{shadow1} for these three models. At first, let us focus on the first model described by Eq.~\eqref{eq:iem1}, and its result is exhibited in the left column of Fig.~\ref{shadow1}. From Fig.~\ref{iemf11}, we observe that the emitted specific intensity peaks abruptly at ISCO and then decays rapidly. The position of ISCO is at $r/M=6$ for the Schwarzschild case ($c_{13}=0$), but for the case of $c_{13}=0.9$ here it changes to $r/M\approx7.4$. It indicates that for the first Einstein-\AE{}ther black hole solution, the increase of the coupling constant $c_{13}$ makes photons around the black hole have larger ISCO. Meanwhile, the observed intensity shown in Fig.~\ref{iobf11} has a peak near $b/M\approx8.5$ corresponding to the direct image. In addition, an extremely narrow peak of comparable intensity appears at $b/M\approx6.2$ corresponding to the lensing ring. Of particular interest, a small peak is present at $b/M\approx6$, resulting by the photon ring. Similar results could be obtained in the Schwarzschild case, with a relatively broad peak on the far right, in turn a narrower, and an extremely narrow peak on the left. These three peaks also correspond to the direct image, the lensing ring, and the photon ring, respectively. However, different from the case of $c_{13}=0.9$, the positions of the three peaks are $b/M\approx7.0$, 5.5, and 5.2~\cite{SEGralla,MGuerrero}. Obviously, the increase of the coupling constant $c_{13}$ makes the starting position of the emission move to the right, and further moves the three peaks of the observed intensity to the right. Next, according to the observed intensity, the two-dimensional image of the Einstein-\AE ther black hole viewed by the observer is given in Fig.~\ref{shadowf11}. From the figure, we find that there is an extremely thin ring in the middle dark region, which is just the lensing ring. Considering that the corresponding transfer function is highly demagnified, the total flux it contributes is very small. Inside the lensing ring, there exists a narrower photon ring. However, it is not noticeable due to the very small observed specific intensity. The contribution of the photon ring can be directly ignored for this set of coupling constants. Moreover, the one outside of the lensing ring is the direct image, which occurs suddenly with a very high brightness and then gradually darkens as it extends outward. Similarly, an observer at the same distance from the black hole would observe a very narrow Schwarzschild lensing ring. But in this case, the radius of the ring is smaller than the one of an Einstein-\AE ther black hole. Besides, there is still a photon ring inside, but the total flux is so small that it can be ignored. The outermost is the direct image from bright to dark, whose radius is still smaller than that of the Einstein-\AE ther black hole. In summary, for the first model that the emission suddenly increases, peaks, and then decays sharply at ISCO, and for the first Einstein-\AE{}ther black hole solution with $c_{123}\neq 0$ and $c_{14}=0$, the increased coupling constant $c_{13}$ allows the observer to see larger black hole images, either the lensing ring or the direct image.

Next, we turn to the second model given in Eq.~\eqref{eq:iem2}, see the middle column of Fig.~\ref{shadow1}. The emission appears suddenly at the photon sphere with $r/M\approx3.8$ in Fig.~\ref{iemf12}. While from Fig.~\ref{iobf12}, we see that the observed intensity is significantly reduced due to the redshift, and the direct image appears at $b/M\approx4.9$. Interestingly, the lensing ring and photon ring are superimposed on the direct image at $b/M\approx6$, forming another peak. It can be seen intuitively from Fig.~\ref{shadowf12}, the brightness suddenly increases at $b/M\approx4.9$, then gradually decreases. At $b/M\approx6$, it increases again, and rapidly dims. Compared to the first model, the maximum observed intensity is reduced here, so that the observer will see a darker image of the black hole. For Schwarzschild black hole, the emission appears suddenly at $r/M=3$ instead of the aforementioned $r/M\approx3.8$ for the case of $c_{13}=0.9$. The corresponding observed intensity still has two peaks. One is the direct image at $b/M\approx3.9$, and the other is the lensing ring and photon ring superimposed on the direct image at $b/M\approx5.2$. Likewise, the redshift effects reduce the observed intensity. Therefore, we find that the effect of the coupling constant $c_{13}$ is still to make the initial position of the emission and the peak positions of the observed intensity move towards the larger value.

We study the third model given by Eq.~\eqref{eq:iem3}, see the right column of Fig.~\ref{shadow1}. In Fig.~\ref{iemf13}, we show the emission from the event horizon ($r/M\approx2.7$) followed by a relatively slowly decay towards the ISCO. The observed intensity in Fig.~\ref{iobf13} is slightly stronger than that in the second model. One could find that the image starts at $b/M\approx4$, which is the position of the event horizon. It indicates that when the emission extends to the event horizon, we could obtain a ``direct" image of it. Besides, a peak appears at $b/M\approx6$, which corresponds to the photon ring. Following it, a second peak, the lensing ring, emerges. The difference between the widths of these two peaks suggests that the lensing ring gives more contributions to the total flux than the photon ring. The black hole image seen by the observer is presented in Fig.~\ref{shadowf13}. Compared to images of the first two models, which consist of two distinct rings, the image of the third model has only one wide ring from dark to bright, then darkened. Besides, this image has greater brightness. For Schwarzschild black hole that is not affected by the coupling constant $c_{13}$, the emission starts at $r/M=2$. And the observed intensity gradually increases from $b/M=3$. Interestingly, unlike the direct superposition of the photon ring and the lensing ring in the second model, here these two rings are superimposed on the direct image, respectively. The first peak appears at $b/M=5.2$, which is the superposition of the photon ring and the direct image. Immediately after the lensing ring is superimposed on the direct image at $b/M=5.4$, then the observed intensity decreases. It is not difficult to find that the role of the coupling constant $c_{13}$ here is still the same as that of the previous two models.

At last, two cases of the second Einstein-\AE{}ther black hole are displayed in Figs.~\ref{shadow2} and \ref{shadow3}, respectively. In Fig.~\ref{shadow2}, we show the case of $c_{123}=0,~c_{13}=0$, and $c_{14}=0.9$. It is worth noting that in this case, it will return to Schwarzschild black hole when $c_{14}$ goes to zero. In the three emission models, the emission suddenly increases at ISCO, the photon sphere, and the event horizon, respectively. Figs.~\ref{iemf21}, \ref{iemf22}, and \ref{iemf23} show the corresponding three starting positions at $r/M=5.3$, 2.7, and 1.7, all of which are obviously smaller than the Schwarzschild case. It just means that the nonvanishing coupling constant $c_{14}$ will make the radius of ISCO, the photon sphere, and the event horizon smaller than that of Schwarzschild black hole. Furthermore, the three images of the observed intensity are similar to the Schwarzschild case. In the first model, there are three independent peaks representing the photon ring, the lensing ring and the direct image, respectively. In the second model, the first peak represents the redshifted direct image, and the second peak is the photon ring and lensing ring superimposed together on the direct image. Two peaks also appear in the third model, but they are the superposition of the photon ring and the lensing ring, respectively, with the direct image. However, compared to the case of Schwarzschild black hole, all peaks are shifted to the left. Figs.~\ref{shadowf21}, \ref{shadowf22}, and \ref{shadowf23} show the two-dimensional images of the Einstein-\AE ther black hole seen by an observer, where each bright ring has a smaller radius than the corresponding Schwarzschild case. Obviously, all these results are caused by the nonvanishing coupling constant $c_{14}$. Another case of $c_{123}=0,~c_{13}=0.5$, and $c_{14}=0$ is shown in Fig.~\ref{shadow3}, where only the coupling constant $c_{13}$ works. Contrary to the role of $c_{14}$, $c_{13}$ makes the radius of ISCO, the photon sphere, and the event horizon larger than that of Schwarzschild black hole, and allows the observer to see a larger image of the black hole. Moreover, all images indicate that the direct image is the main contributor to the total flux, and the lensing ring could only make some very small contribution, whereas the role of the photon ring is negligible. Besides, for each model, we observe that the optical appearance significantly depends on these coupling constants.

\section{Conclusions}
 \label{conclusions}

In this paper, we have investigated the motion of photons around a static and spherically symmetric black hole in Einstein-\AE ther theory, and studied the optical appearance of the black hole surrounded by an optically thin and geometrically thin accretion disk.

Firstly, we briefly reviewed the Einstein-\AE ther theory~\cite{TJacobson1}, which breaks the Lorentz symmetry locally by introducing a dynamical, unit time-like \AE ther field $u^\mu$. Static and spherically symmetric black hole solutions are found in this theory, and the solutions have two exact forms $f_1(r)$ and $f_2(r)$ while with different coupling constants~\cite{CEling2,EBarausse1}.

Next, we studied the motion of photons in this interesting black hole backgrounds. From the Euler-Lagrange equation, we obtained the specific expression of the radial component by introducing two conserved quantities, the energy $E$ and angular momentum $L$. Then, the radius of the photon sphere $r_p$ and the corresponding critical impact parameter $b_c$ in different black hole solutions were acquired from the conditions for circular orbits. For photons with different impact parameters, we also calculated the total deflection angle $\phi$. In addition, following the novel viewpoints proposed by Gralla, Holz, and Wald~\cite{SEGralla}, we studied the behavior of photons with different total number $n$ of orbits. The behavior of photons as a function of the impact parameter $b/M$ is shown in Figs.~\ref{fig1} and \ref{fig2}. On the one hand, for the first Einstein-\AE{}ther black hole solution, both the radii of the event horizon and photon sphere increase with the coupling constant $c_{13}$. Thus the observers at infinity will see a larger internal dark region and a narrower lensing ring. On the other hand, for the second Einstein-\AE{}ther black hole solution, there are two characteristic cases: (a) Coupling constant $c_{13}$ is fixed. It is found that the radii of the event horizon and photon sphere decrease with $c_{14}$, leading to a smaller internal dark area, but a wider lensing ring; (b) Coupling constant $c_{14}$ is fixed. The results are reverse with the increase of $c_{13}$. Furthermore, we compared the behavior of photons in Einstein-\AE ther theory with that in GR.

Finally, we delineated the images of the black holes surrounded by a thin disk emission. Three toy-model functions were used to study the specific intensity of emission, starting from ISCO, the photon sphere, and the event horizon, respectively. To obtain the observed specific intensity, we discussed its relationship to the emitted intensity. With the ray-tracing procedure, photons could get extra brightness when they intersect the disk. The more intersections, the greater the brightness obtained. Hence we investigated the transfer function $r_m(b)$ for the first three transfer functions with different coupling constants in Fig.~\ref{fig3}.

The slope $\text{d}r/\text{d}b$ of the transfer function is the demagnification factor, which characterizes the demagnified scale of the transfer function. The first transfer function corresponding to the direct image has a very flat slope, which indicates that the image profile is just a gravitational redshifted source profile. The second transfer function is for the lensing ring, where its large slope will give a high demagnification image. The third one corresponds to the photon ring, whose extremely demagnified slope implies that the brightness could be very large. Consequently, the contributions of the photon ring and the further images to the total flux are negligible. Basing on above discussions, we investigated the optical appearance of the black holes. We found that the image is highly dependent on the emitted model, as well as the coupling constants. Moreover, all three emitted models suggest that the direct image dominates in the total flux, and the lensing ring is a bright but extremely narrow ring that cannot be easily observed, while the photon ring contributes negligibly and is almost invisible.

\section*{Acknowledgements}
This work was supported by the National Natural Science Foundation of China (Grants No. 12075103, No. 11675064, and No. 12047501), the 111 Project (Grant No. B20063), and the Fundamental Research Funds for the Central Universities (No. Lzujbky-2019-ct06).

\end{document}